\documentclass[a4paper,fleqn]{cas-sc}

\usepackage[T1]{fontenc}
\usepackage[polutonikogreek,italian,english]{babel}
\usepackage[page,toc,titletoc,title]{appendix}
\usepackage[numbers,sort&compress]{natbib}
\usepackage{graphicx}
\usepackage{graphics}
\usepackage{braket}
\usepackage{bbold}
\usepackage{amssymb}
\usepackage{amsmath}
\usepackage{nicefrac}
\usepackage{dcolumn}% Align table columns on decimal point
\usepackage{bm}% bold math
\usepackage{slashed}
\usepackage{datetime}
\usepackage{mciteplus}
\usepackage{multirow}
\usepackage{bbold}
\usepackage{color}
\usepackage{xcolor}
\usepackage{float}
\usepackage[utf8]{inputenc}
\usepackage[normalem]{ulem}
\usepackage{acro}

\DeclareAcronym{AI}{
  short=AI,
  long=Artificial Intelligence,
}
\DeclareAcronym{BSM}{
  short=BSM,
  long=Beyond-the-Standard-Model,
}
\DeclareAcronym{QCD}{
  short=QCD,
  long=Quantum ChromoDynamics,
}
\DeclareAcronym{SM}{
  short=SM,
  long=Standard Model,
}
\DeclareAcronym{CP}{
  short=CP,
  long=Charge-Parity,
}
\DeclareAcronym{SLAC}{
  short=SLAC,
  long=Stanford Linear Accelerator Center,
}
\DeclareAcronym{BNL}{
  short=BNL,
  long=Brookhaven National Laboratory,
}
\DeclareAcronym{FCNCs}{
  short=FCNCs,
  long=Flavor-Changing Neutral Currents,
}
\DeclareAcronym{GIM}{
  short=GIM,
  long=Glashow--Iliopoulos--Maiani,
}
\DeclareAcronym{CEM}{
  short=CEM,
  long= Color Evaporation Model,
}
\DeclareAcronym{CSM}{
  short=CSM,
  long= Color Singlet Mechanism,
}
\DeclareAcronym{CO}{
  short=CO,
  long=Color Octet,
}
\DeclareAcronym{NRQCD}{
  short=NRQCD,
  long=NonRelativistic QCD,
}
\DeclareAcronym{LDME}{
  short=LDME,
  long=Long-Distance Matrix Element,
}
\DeclareAcronym{LO}{
  short=LO,
  long=Leading Order,
}
\DeclareAcronym{NLO}{
  short=NLO,
  long=Next-to-Leading Order,
}
\DeclareAcronym{NNLO}{
  short=NNLO,
  long=Next-to-NLO,
}
\DeclareAcronym{MHOUs}{
  short=MHOUs,
  long=Missing Higher-Order Uncertainties,
}
\DeclareAcronym{DY}{
  short=DY,
  long=Drell--Yan,
}
\DeclareAcronym{DIS}{
  short=DIS,
  long=Deep-Inelastic Scattering,
}
\DeclareAcronym{SIDIS}{
  short=SIDIS,
  long=Semi-Inclusive Deep-Inelastic Scattering,
}
\DeclareAcronym{SIA}{
  short=SIA,
  long=Single-Inclusive Annihilation,
}
\DeclareAcronym{DGLAP}{
  short=DGLAP,
  long=Dokshitzer--Gribov--Lipatov--Altarelli--Parisi,
}
\DeclareAcronym{PDFs}{
  short=PDFs,
  long=Parton Distribution Functions,
}
\DeclareAcronym{FFs}{
  short=FFs,
  long=Fragmentation Functions,
}
\DeclareAcronym{MPIs}{
  short=MPIs,
  long=Multi-Parton Interactions,
}
\DeclareAcronym{DPS}{
  short=DPS,
  long=Double-Parton Scattering,
}
\DeclareAcronym{SCET}{
  short=SCET,
  long=Soft and Collinear Effective Theory,
}
\DeclareAcronym{TM}{
  short=TM,
  long=Transverse-Momentum,
}
\DeclareAcronym{TMD}{
  short=TMD,
  long=Transverse-Momentum-Dependent,
}
\DeclareAcronym{FFNS}{
  short=FFNS,
  long=Fixed-Flavor Number-Scheme,
}
\DeclareAcronym{VFNS}{
  short=VFNS,
  long=Variable-Flavor Number-Scheme,
}
\DeclareAcronym{GM-VFNS}{
  short=GM-VFNS,
  long=General-Mass Variable-Flavor Number-Scheme,
}
\DeclareAcronym{ABF}{
  short=ABF,
  long=Altarelli--Ball--Forte,
}
\DeclareAcronym{BFKL}{
  short=BFKL,
  long=Balitsky--Fadin--Kuraev--Lipatov,
}
\DeclareAcronym{LL}{
  short=LL,
  long=Leading Logarithmic,
}
\DeclareAcronym{NLL}{
  short=NLL,
  long=Next-to-Leading Logarithmic,
}
\DeclareAcronym{NNLL}{
  short=NNLL,
  long=Next-to-NLL,
}
\DeclareAcronym{LVM}{
  short=LVM,
  long=Light Vector Meson,
}
\DeclareAcronym{UGD}{
  short=UGD,
  long=Unintegrated Gluon Distribution,
}
\DeclareAcronym{GPDs}{
  short=GPD,
  long=Generalized Parton Densities,
}
\DeclareAcronym{LHC}{
  short=LHC,
  long=Large Hadron Collider,
}
\DeclareAcronym{EIC}{
  short=EIC,
  long=Electron-Ion Collider,
}
\DeclareAcronym{FCC}{
  short=FCC,
  long=Future Circular Collider,
}
\DeclareAcronym{FPFs}{
  short=FPFs,
  long=Forward Physics Facilities,
}
\DeclareAcronym{HFAG}{
  short=HFAG,
  long=Heavy Flavor Averaging Group,
}
\DeclareAcronym{SCA}{
  short=SCA,
  long=Small-Cone Algorithm
}
\DeclareAcronym{BLM}{
  short=BLM,
  long=Brodsky--Lepage--Mackenzie,
}
\DeclareAcronym{SNAJ}{
  short=SNAJ,
  long=Suzuki--Nejad--Amiri--Ji,
}
\DeclareAcronym{MSb}{
  short={$\boldsymbol{\overline{\rm MS}}$},
  long=Modified Minimal Subtraction,
}
\DeclareAcronym{MOM}{
  short=MOM,
  long=MOMentum,
}
\DeclareAcronym{HELL}{
  short={HELL},
  long=High Energy Large Logarithms,
}

\newcommand{\deffont}[1]{\begin{otherlanguage*}{polutonikogreek}#1\end{otherlanguage*}}

\def\tsc#1{\csdef{#1}{\textsc{\lowercase{#1}}\xspace}}
\tsc{WGM}
\tsc{QE}
\newcommand{\drv}{{\rm d}}

\newcommand{\as}{\alpha_s}
\newcommand{\LQCD}{\Lambda_{\rm QCD}}
\newcommand{\MSb}{\overline{\rm MS}}

\newcommand{\LL}{{\rm LL/LO}}
\newcommand{\NLL}{{\rm NLL/NLO}}
\newcommand{\NLLp}{{\rm NLL/NLO^+}}

\newcommand{\HENLOp}{{\rm HE}\mbox{-}{\rm NLO^+}}

\newcommand{\CnLL}{{\cal C}_n^\LL}

\newcommand{\CnNLLp}{{\cal C}_n^\NLLp}

\newcommand{\CnHENLOp}{{\cal C}_n^{{\rm HE}\text{-}{\rm NLO}^+}}
\newcommand{\DY}{\Delta Y}

\newcommand{\F}{{\cal F}}
\newcommand{\J}{{\cal J}}

\newcommand{\Hb}{{\cal H}_b}

\newcommand{\XQq}{ X_{Qq\bar{Q}\bar{q}}}

\newcommand{{\Jethad}}{\textsc{Jethad}}
\newcommand{{\symJethad}}{\tt symJETHAD}
\newcommand{{\Hell}}{\tt HELL}
\newcommand{{\RadISH}}{\textsc{RadISH}}

\begin{document}
\let\WriteBookmarks\relax
\def\floatpagepagefraction{1}
\def\textpagefraction{.001}

\shorttitle{Forward \& Far-Forward Heavy Hadrons with {\Jethad}: A High-energy Viewpoint}    

\shortauthors{Celiberto, Francesco Giovanni}  

\title []{\Huge Forward \& Far-Forward Heavy Hadrons with {\Jethad}: A High-energy Viewpoint}  

\author[1]{Francesco Giovanni Celiberto}[orcid=0000-0003-3299-2203]

\cormark[1]

%\fnmark[<footnote mark no>]

\ead{francesco.celiberto@uah.es}

%\ead[url]{<URL>}

\affiliation[1]{organization={Universidad de Alcal\'a (UAH), Departamento de F\'isica y Matem\'aticas},
            addressline={Campus Universitario}, 
            city={Alcal\'a de Henares},
%          citysep={}, % Uncomment if no comma needed between city and postcode
            postcode={E-28805}, 
            state={Madrid},
            country={Spain}}

%\cortext[1]{Corresponding author}



%-----------------------------------------
\begin{abstract}
Inspired by recent findings that semi-inclusive detections of heavy hadrons exhibit fair stabilization patterns in high-energy resummed distributions against (missing) higher-order corrections, we review and extend our studies on the hadroproduction of light and heavy hadrons tagged in forward and far-forward rapidity ranges.
We analyze the $\NLLp$ behavior of rapidity rates and angular multiplicities via the {\Jethad} method, where the resummation of next-to-leading energy logarithms and beyond is consistently embodied in the collinear picture.
We explore kinematic regions that are within LHC typical acceptances, as well as novel sectors accessible thanks the combined tagging of a far-forward light or heavy hadron at future Forward Physics Facilities and a of central particle at LHC experiments via a precise timing-coincidence setup.

\end{abstract}
%-----------------------------------------


%\item 
%\item 
%\item 

% Keywords
\begin{keywords}
 Heavy Flavor \sep
 QCD Resummation \sep
 LHC Phenomenology \sep 
 Forward Physics Facilities \sep
 Hadron detections \sep
\end{keywords}

\maketitle

\newcounter{appcnt}

%%%%%%%%%%%%%%%%%%%%%%%%%%%%%%%%%%%%%%%%%%
%%%\setcounter{section}{-1} %% Remove this when starting to work on the template.

\tableofcontents
\clearpage
%\printacronyms

\setlength{\parskip}{3pt}%
%\setlength{\parindent}{0pt}%

%-----------------------------------------
\section{Introduction}
\label{sec:introduction}
%-----------------------------------------

New paths in the exploration of fundamental interactions by next-generation colliders~\cite{Anchordoqui:2021ghd,Feng:2022inv,Hentschinski:2022xnd,Accardi:2012qut,AbdulKhalek:2021gbh,Khalek:2022bzd,Acosta:2022ejc,AlexanderAryshev:2022pkx,Brunner:2022usy,Arbuzov:2020cqg,Abazov:2021hku,Bernardi:2022hny,Chapon:2020heu,Amoroso:2022eow,Celiberto:2018hdy,Klein:2020nvu,2064676,MuonCollider:2022xlm,Aime:2022flm,MuonCollider:2022ded,Accettura:2023ked,Vignaroli:2023rxr,Black:2022cth,Dawson:2022zbb,Bose:2022obr,Begel:2022kwp,Abir:2023fpo,Accardi:2023chb,Mangano:2016jyj,FCC:2018byv,FCC:2018evy,FCC:2018vvp,FCC:2018bvk} mark the turn of a new era in particle physics. Venturing into uncharted kinematic territories permits stringent analyses of the \ac{SM} and direct or indirect quests for deviations from its predictions.
The strong-interaction sector presents significant challenges within the SM. Here, the interplay between perturbative and nonperturbative aspects of \ac{QCD} gives rise to unresolved puzzles regarding fundamental questions, including the origin of hadron mass and spin, as well as the behavior of QCD observables in critical kinematic regimes.

Precise examinations of the dynamics underlying strong interactions rely on two essential components. Firstly, the ability to perform increasingly accurate calculations of high-energy parton scatterings through higher-order perturbative QCD techniques. Secondly, the understanding of proton structure, dictated by the motion and spin interactions among constituent partons.
A series of successes in describing data for hadron, lepton, and lepton-hadron reactions have been achieved through collinear factorization~\cite{Collins:1989gx,Sterman:1995fz}, where partonic cross sections, computed within pure perturbative QCD, are convoluted with collinear \ac{PDFs}.

These PDFs encode information about the likelihood of finding a parton inside the struck hadron with a specific longitudinal momentum fraction, $x$. They evolve according to the \ac{DGLAP} equation~\cite{Gribov:1972ri,Gribov:1972rt,Lipatov:1974qm,Altarelli:1977zs,Dokshitzer:1977sg}. Collinear PDFs are well-suited for describing inclusive or semi-inclusive observables weakly sensitive to low-transverse momentum regimes.
Similarly, collinear \ac{FFs} portray the production mechanism of identified hadrons, detailing the probability of generating a specific final-state hadron with momentum fraction $z$ from an outgoing collinear parton with longitudinal fraction, $\zeta \equiv x/z$.

However, by relying on collinear PDFs only, one overlooks information about the transverse-space distribution and motion of partons. Thus, the description provided by collinear factorization can be seen as a one-dimensional mapping of hadron properties.
Achieving an accurate portrayal of low-transverse-momentum observables necessitates embracing a three-dimensional perspective, which permits the capture of intrinsic effects stemming from the transverse motion and spin of partons, and their interaction with the polarization state of the parent hadron. Such a tomographic representation of hadrons is naturally provided by the \ac{TMD} factorization (see Refs.~\cite{Collins:1981uk,Collins:2011zzd} and reference therein).

Furthermore, given the nonperturbative nature of parton densities and fragmentation functions, they must be extracted from data through \emph{global fits} encompassing various hadronic processes.
Despite the significant successes of the collinear approach, certain kinematic regions demand a departure from the fixed-order, DGLAP-driven description. Enhanced logarithmic contributions, which enter the perturbative expansion of the strong running coupling $\alpha_s$ with increasingly higher powers, must be included to all orders to restore the convergence of the perturbative series. These logarithms, depending on the kinematic regimes, necessitate specific all-order \emph{resummation} techniques.

For instance, accurately describing the differential distributions for inclusive hadron, boson, or \ac{DY} lepton production at low $|\vec q_T|$ demands employing \ac{TM} resummation techniques~\cite{Catani:2000vq,Bozzi:2005wk,Bozzi:2008bb,Catani:2010pd,Catani:2011kr,Catani:2013tia,Catani:2015vma,Duhr:2022yyp}.
TM-resummed predictions have been recently extended to various reactions, including hadroproduction of photons~\cite{Cieri:2015rqa,Alioli:2020qrd,Becher:2020ugp,Neumann:2021zkb}, Higgs bosons~\cite{Ferrera:2016prr}, $W$-boson pairs~\cite{Ju:2021lah}, boson-jet~\cite{Monni:2019yyr,Buonocore:2021akg}, $Z$-photon systems~\cite{Wiesemann:2020gbm}, as well as DY and Higgs spectra~\cite{Ebert:2020dfc,Re:2021con,Chen:2022cgv,Neumann:2022lft,Bizon:2017rah,Billis:2021ecs,Re:2021con,Caola:2022ayt}.
Moreover, almost back-to-back final states lead to Sudakov-type logarithms, which should be resummed as well~\cite{Mueller:2012uf,Mueller:2013wwa,Marzani:2015oyb,Mueller:2015ael,Xiao:2018esv}.

Conversely, when a physical observable is defined and/or measured close to its phase space edges, Sudakov effects originating from soft and collinear gluon emissions near \emph{threshold} become significant and require appropriate resummation techniques. Various approaches exist in the literature for achieving threshold resummation for inclusive rates~\cite{Sterman:1986aj,Catani:1989ne,Catani:1996yz,Bonciani:2003nt,deFlorian:2005fzc,Ahrens:2009cxz,deFlorian:2012yg,Forte:2021wxe,Mukherjee:2006uu,Bolzoni:2006ky,Becher:2006nr,Becher:2007ty,Bonvini:2010tp,Bonvini:2018ixe,Ahmed:2014era,Banerjee:2018vvb,Duhr:2022cob,Shi:2021hwx,Wang:2022zdu}. Recent studies have also addressed resummation for rapidity distributions~\cite{Mukherjee:2006uu,Bolzoni:2006ky,Becher:2006nr,Becher:2007ty,Bonvini:2010tp,Ahmed:2014era,Banerjee:2018vvb}.

The standard fixed-order description of cross sections has been enhanced by incorporating threshold resummation for numerous processes, including \ac{DY}~\cite{Moch:2005ky,Idilbi:2006dg,Catani:2014uta,Ajjath:2020rci,Ajjath:2020lwb,Ajjath:2021pre,Ahmed:2020nci,Ajjath:2021lvg}, scalar and pseudo-scalar Higgs production~\cite{Kramer:1996iq,Catani:2003zt,Moch:2005ky,Bonvini:2012an,Bonvini:2014joa,Catani:2014uta,Bonvini:2014tea,Bonvini:2016frm,Beneke:2019mua,Ajjath:2020sjk,Ajjath:2020lwb,Ahmed:2020nci,Ajjath:2021bbm,deFlorian:2007sr,Schmidt:2015cea,Ahmed:2016otz,Bhattacharya:2019oun,Bhattacharya:2021hae}, bottom annihilation~\cite{Bonvini:2016fgf,Ajjath:2019neu,Ahmed:2020nci}, \ac{DIS}~\cite{Moch:2005ba,Das:2019btv,Ajjath:2020sjk,Abele:2022wuy}, electron-positron \ac{SIA}~\cite{Ajjath:2020sjk}, and spin-2 boson production~\cite{Das:2019bxi,Das:2020gie}. Additionally, a combined TM-plus-threshold resummation for TM distributions of colorless final states has been developed~\cite{Muselli:2017bad}.

From a partonic perspective, the threshold regime corresponds to the large-$x$ limit. Prime determinations of large-$x$ improved collinear PDFs were achieved in Ref.~\cite{Bonvini:2015ira}. 
Additionally, as $x$ approaches one, the impact of \emph{target-mass} power corrections becomes significant. These corrections have been studied extensively in the literature~\cite{Nachtmann:1973mr,Georgi:1976ve,Barbieri:1976rd,Ellis:1982wd,Ellis:1982cd,Schienbein:2007gr,Accardi:2008ne,Accardi:2008pc,Accardi:2013pra}, particularly in the context of large-$x$ DIS events~\cite{Accardi:2014qda}.

Another crucial kinematic regime, sensitive to logarithmic enhancements, is the \emph{semi-hard} (or Regge--Gribov) sector~\cite{Gribov:1983ivg,Celiberto:2017ius,Bolognino:2021bjd,Mohammed:2022gbk}, characterized by the scale hierarchy $\sqrt{s} \gg \{Q\} \ll \LQCD$, with $\sqrt{s}$ being the center-of-mass energy, $\{Q\}$ one or a set of process-characteristic energy scales, and $\LQCD$ the QCD hadronization scale.
Here, large energy logarithms of the form $\ln (s/Q^2)$ compensate for the narrowness of $\alpha_s$, potentially spoiling the perturbative-series convergence.

Like in previous cases, these logarithms must be resummed to all orders. 
The \ac{BFKL} formalism is the most powerful tool for such resummations~\cite{Fadin:1975cb,Kuraev:1976ge,Kuraev:1977fs,Balitsky:1978ic,Fadin:1998sh}. 
BFKL resummation accounts for contributions proportional to $\alpha_s^n \ln (s/Q^2)^n$ up to the \ac{LL} level, and terms proportional to $\alpha_s^{n+1} \ln (s/Q^2)^n$ at the \ac{NLL} level.

In the BFKL framework, the imaginary part of amplitudes (and also cross sections of inclusive processes, via to the \emph{optical theorem}~\cite{optical_theorem_Newton}) takes the form of a high-energy factorization, where TM-dependent functions play a key role~\cite{Catani:1990xk,Catani:1990eg,Catani:1993ww}. Specifically, the BFKL amplitude is a convolution between two singly off-shell emission functions (also known in the BFKL jargon as forward-production \emph{impact factors}), describing the transition from each parent particle to the outgoing objects in its fragmentation region, and a Green's function, which evolves according to the BFKL integral equation. 
The kernel of this equation has been computed with \ac{NLO} accuracy for forward scatterings~\cite{Fadin:1998py,Ciafaloni:1998gs,Fadin:1998jv,Fadin:2000kx,Fadin:2000hu,Fadin:2004zq,Fadin:2005zj} and partially at next-to-NLO~\cite{Caola:2021izf,Falcioni:2021dgr,DelDuca:2021vjq,Byrne:2022wzk,Fadin:2023roz,Byrne:2023nqx}.

Conversely, emission functions are sensitive to the given final state. They are also $\{Q\}$-dependent, but $s$-independent.
Thus, they represent the most intricate piece of a BFKL computation, and only few of them are currently known at NLO.
We mention: 
a) collinear-parton functions~\cite{Fadin:1999de,Fadin:1999df}, which are needed to compute 
b) forward-jet~\cite{Bartels:2001ge,Bartels:2002yj,Caporale:2011cc,Caporale:2012ih,Ivanov:2012ms,Colferai:2015zfa} and 
c) light-hadron~\cite{Ivanov:2012iv},
d) virtual photon to light vector meson~\cite{Ivanov:2004pp}, 
e) light-by-light~\cite{Bartels:2000gt,Bartels:2001mv,Bartels:2002uz,Bartels:2004bi,Fadin:2001ap,Balitsky:2012bs}, and
f) forward-Higgs boson in the infinite top-mass limit~\cite{Hentschinski:2020tbi,Celiberto:2022fgx} (see also Refs.~\cite{Hentschinski:2022sko,Fucilla:2022whr}) emission functions.
Considering the \ac{LO} only, one has
DY pairs~\cite{Hentschinski:2012poz,Motyka:2014lya},
heavy-quark pairs~\cite{Celiberto:2017nyx,Bolognino:2019ccd,Bolognino:2019yls}, and low-$|\vec q_T|$ $J/\psi$\cite{Boussarie:2017oae,Boussarie:2015jar,Boussarie:2016gaq,Boussarie:2017xdy} impact factors.

A primary category of processes that act as probe channels for high-energy resummation involves single forward emissions. In these cases, the cross section for an inclusive process follows the typical BFKL-factorized structure. Specifically, when at least one hadron participates in the initial state, the impact factor governing the production of the forward identified particle is convoluted with the BFKL Green's function and a nonperturbative quantity known as the hadron impact factor. 
The sub-convolution of the last two components provides an operational definition of the BFKL \ac{UGD}. 

The hadron impact factor serves as the initial-scale UGD, while the Green's function governs its low-$x$ evolution. Given the forward kinematics, the struck parton is predominantly a gluon with a small longitudinal-momentum fraction. Hence, in this context, high-energy resummation effectively amounts to low-$x$ resummation.
A similar high-energy factorization formula applies to the imaginary part of the amplitude for exclusive single forward processes. This is feasible because, in the forward limit, skewness effects are suppressed, allowing for the use of the same UGD. In more general off-forward configurations, one would consider low-$x$ enhanced GPDs~\cite{Diehl:2003ny,Diehl:2015uka,Muller:1994ses,Belitsky:2005qn}.

An intriguing subset of inclusive forward reactions involves proton-initiated processes. Here, a hybrid high-energy and collinear factorization is employed, where the forward object originates from a fast parton with a moderate $x$, described by a collinear PDF, while the other proton is characterized by the UGD.
Studies on the BFKL UGD trace back to the growing interest in forward physics at HERA. Investigations of DIS structure functions at low $x$ were undertaken in Refs.~\cite{Hentschinski:2012kr,Hentschinski:2013id}. 

Subsequently, different UGD models were scrutinized against HERA data for exclusive light vector-meson electroproduction in references~\cite{Anikin:2011sa,Besse:2013muy,Bolognino:2018rhb,Bolognino:2018mlw,Bolognino:2019bko,Bolognino:2019pba,Celiberto:2019slj}.
Evidence of low-$x$ dynamics is expected to emerge from $\rho$-meson studies at the \ac{EIC}~\cite{Bolognino:2021niq,Bolognino:2021gjm,Bolognino:2022uty,Celiberto:2022fam,Bolognino:2022ndh}. 
Similarly, insights are sought from the photoemission of quarkonium states~\cite{Bautista:2016xnp,Garcia:2019tne,Hentschinski:2020yfm,Peredo:2023oym,GayDucati:2013sss,GayDucati:2016ryh,Goncalves:2017wgg,Goncalves:2018blz,Cepila:2017nef,Guzey:2020ntc,Jenkovszky:2021sis,Flore:2020jau,ColpaniSerri:2021bla}.
Forward DY and single inclusive $b$-quark tags at the \ac{LHC} serve as hadronic probes for the UGD~\cite{Motyka:2014lya,Brzeminski:2016lwh,Motyka:2016lta,Celiberto:2018muu,Chachamis:2015ona}.

Besides single forward emissions, the low-$x$ QCD sector can also be probed via gluon-induced single central productions. In these processes, the cross sections are formulated in a pure high-energy factorized form, involving a convolution between two BFKL UGDs and a central-production impact factor, embodying a doubly off-shell coefficient function.
Due to its doubly off-shell nature (involving virtual gluons $g^*g^*$), with gluon virtualities determined by their transverse momenta, computing the coefficient function is considerably more intricate compared to the forward-case scenarios. To our knowledge, the coefficient function  depicting the inclusive emission of a central, light-flavored jet is the only one known at NLO~\cite{Bartels:2006hg}.

A very powerful formalism aimed at enhancing standard fixed-order computation of central processes via the low-$x$ resummation is the \ac{ABF} scheme~\cite{Ball:1995vc,Ball:1997vf,Altarelli:2001ji,Altarelli:2003hk,Altarelli:2005ni,Altarelli:2008aj,White:2006yh}, where $\kappa_T$-factorization theorems~\cite{Catani:1990xk,Catani:1990eg,Collins:1991ty,Catani:1993ww,Ball:2007ra} permits to combine DGLAP and BFKL inputs. 
The high-energy series is stabilized by imposing consistency conditions based on duality aspects, symmetrizing the BFKL kernel in (anti-)collinear regions of the phase space, and encoding such contributions to the running coupling.

Significant progress has been achieved in small-$x$ studies within the ABF formalism, particularly in the context of inclusive central emissions. Notable advancements include investigations into the inclusive central emissions of Higgs bosons in gluon-gluon fusion~\cite{Marzani:2008az,Caola:2010kv,Caola:2011wq,Forte:2015gve} and higher-order corrections to top-quark pair emissions~\cite{Muselli:2015kba}. 
Applications ABF to resummed inclusive or differential distributions for Higgs-boson and heavy-flavor hadroproductions were done by means of the \ac{HELL} method~\cite{Bonvini:2018ixe,Silvetti:2022hyc,Silvetti:2023suu}.
Moreover, this framework has been instrumental in extracting low-$x$ enhanced collinear PDFs for the first time~\cite{Ball:2017otu,Abdolmaleki:2018jln,Bonvini:2019wxf}. Subsequently, the information from these collinear PDFs was utilized to constrain the parameters of initial-scale unpolarized and helicity gluon TMD PDFs~\cite{Bacchetta:2020vty}.

Another class of processes serving as promising channel whereby unveiling the onset of high-energy QCD dynamics, is represented by inclusive hadroproductions of two particles emitted with transverse momenta well above $\LQCD$ and strongly separated in rapidity.\footnote{High-energy effects were also seen in photon-initiated processes, such as the ($\gamma^*\gamma^*$) reaction~\cite{Brodsky:2002ka,Chirilli:2014dcb,Ivanov:2014hpa}, the exclusive di-meson leptoproduction~\cite{Segond:2007fj,Ivanov:2005gn,Ivanov:2006gt}, and the semi-inclusive heavy-quark pair photodetection~\cite{Celiberto:2017nyx}.}
Contrarily to single forward and central processes, forward-plus-backward two-particle hadroproduction rates are sensitive to enhanced energy logarithms even at moderate values of $x$, due to the peculiar kinematic ranges in transverse momentum and rapidity currently covered by acceptances of LHC detectors.

Thus, on the one hand, a collinear treatment still holds here.
On the other hand, however, high rapidity intervals ($\Delta Y$) correspond to large TM exchanges in the $t$-channel, leading to the emergence of energy logarithms. 
In such cases, a high-energy factorization treatment is needed, and it is inherently provided by the BFKL formalism. Consequently, another form of hybrid high-energy and collinear factorization is established~\cite{Colferai:2010wu,Celiberto:2020wpk,Celiberto:2020tmb,Bolognino:2021mrc,Celiberto:2022rfj,Celiberto:2022dyf,Celiberto:2023fzz} (see Refs.~\cite{Deak:2009xt,vanHameren:2015uia,Deak:2018obv,VanHaevermaet:2020rro,vanHameren:2022mtk,Giachino:2023loc,Guiot:2024oja} for a close-in-spirit formalism), wherein high-energy resummed partonic cross sections are derived directly from BFKL and subsequently convoluted with collinear PDFs. This approach allows for a comprehensive description of processes occurring in kinematic regimes characterized by both large rapidity intervals and transverse momenta.

The ``mother'' reaction of forward-plus-backward inclusive hadroproductions is the Mueller--Navelet~\cite{Mueller:1986ey} tagging of two light jets at large momenta and $\DY$, for which a series of phenomenological studies have appeared so far~\cite{Marquet:2007xx,Colferai:2010wu,Caporale:2012ih,Ducloue:2013hia,Ducloue:2013bva,Caporale:2013uva,Caporale:2014gpa,Ducloue:2015jba,Celiberto:2015yba,Celiberto:2015mpa,Caporale:2015uva,Mueller:2015ael,Celiberto:2016ygs,Celiberto:2016vva,Caporale:2018qnm,deLeon:2021ecb,Celiberto:2022gji,Egorov:2023duz} and they have been compared with CMS data at $\sqrt{s} = 7\mbox{ TeV}$~\cite{Khachatryan:2016udy,CMS:2021maw}. 
Exploring further observables that are sensitive to more exclusive final states offers additional avenues for uncovering clues about the onset of BFKL dynamics. These channels complement the insights provided by Mueller--Navelet channels and offer a deeper understanding of the underlying processes. 

By studying such observables, we can gain a more comprehensive view of how BFKL dynamics manifest across various exclusive final states, thereby enriching our understanding of high-energy QCD phenomena.
We can mention: light-flavored di-hadron~\cite{Celiberto:2016hae,Celiberto:2016zgb,Celiberto:2017ptm,Celiberto:2017uae,Celiberto:2017ydk}, hadron-plus-jet~\cite{Bolognino:2018oth,Bolognino:2019cac,Bolognino:2019yqj,Celiberto:2020wpk,Celiberto:2020rxb,Celiberto:2022kxx}, multi-jet~\cite{Caporale:2015vya,Caporale:2015int,Caporale:2016soq,Caporale:2016vxt,Caporale:2016xku,Celiberto:2016vhn,Caporale:2016djm,Caporale:2016pqe,Chachamis:2016qct,Chachamis:2016lyi,Caporale:2016lnh,Caporale:2016zkc,Caporale:2017jqj,Chachamis:2017vfa} and Drell--Yan~\cite{Golec-Biernat:2018kem} angular distributions, Higgs-jet rapidity and transverse-momentum distributions~\cite{Celiberto:2020tmb,Celiberto:2021fjf,Celiberto:2021tky,Celiberto:2021txb,Celiberto:2021xpm}, heavy-flavored jet~\cite{Bolognino:2021mrc,Bolognino:2021hxx} and hadron~\cite{Boussarie:2017oae,Celiberto:2017nyx,Bolognino:2019ouc,Bolognino:2019yls,Bolognino:2019ccd,Celiberto:2021dzy,Celiberto:2021fdp,Bolognino:2022wgl,Celiberto:2022dyf,Celiberto:2022grc,Bolognino:2022paj,Celiberto:2022keu,Celiberto:2022zdg,Celiberto:2022kza,Celiberto:2024omj,Celiberto:2023rzw,Celiberto:2024mrq,Celiberto:2024mab} distributions.

Analyses on angular correlations for light-jet and/or light-hadron detections have played a crucial role in distinguishing between high-energy resummed and fixed-order calculations. By employing asymmetric TM ranges, we have been able to decisively discriminate between these approaches, shedding light on the underlying dynamics of high-energy QCD processes~\cite{Celiberto:2015yba,Celiberto:2015mpa,Celiberto:2020wpk}.
However, these studies have also revealed significant challenges associated with higher-order BFKL corrections. 

Specifically, NLL contributions have been found to be comparable in magnitude to LL terms but with opposite signs, leading to instability in the high-energy series. This sensitivity to the variations of renormalization ($\mu_R$) and factorization ($\mu_F$) scales, aimed at gauging the size of \ac{MHOUs}, poses a significant obstacle to achieving reliable theoretical predictions.

Efforts to address these challenges have included the adoption of scale-optimization methods such as the \ac{BLM} optimization~\cite{Brodsky:1996sg,Brodsky:1997sd,Brodsky:1998kn,Brodsky:2002ka} in its semi-hard oriented version~\cite{Caporale:2015uva}. While this approach has shown some success in partially mitigating instabilities in azimuthal correlations, it has proven ineffective for cross sections. In particular, the optimal scales obtained using this method have often been much larger than the natural scales dictated by kinematics~\cite{Celiberto:2020wpk}, resulting in a substantial and unphysical reduction in statistical precision. As a result, achieving precision in the study of inclusive forward-backward light-flavored objects has remained elusive despite these efforts.

Recent studies have provided corroborating indication of a stabilization of the high-energy resummation under higher-order corrections and MHOUs studies, particularly in the context of semi-hard Higgs-boson inclusive emissions~\cite{Celiberto:2020tmb,Mohammed:2022gbk,Celiberto:2023rtu,Celiberto:2023uuk,Celiberto:2023eba,Celiberto:2023nym,Celiberto:2023dkr,Celiberto:2023rqp,Celiberto:2022qbh}. This stabilization trend has been observed also in analyses focusing on semi-inclusive emissions of $\Lambda_c$ baryons~\cite{Celiberto:2021dzy} or singly bottomed hadrons at the LHC~\cite{Celiberto:2021fdp}.

A key observation is a clear signature of stabilization, which is directly linked to the distinctive behavior exhibited by \ac{VFNS}~\cite{Mele:1990cw,Cacciari:1993mq,Buza:1996wv} collinear FFs governing the production of these singly heavy-flavored particles at high transverse momentum. These findings mark a significant advancement in our understanding of high-energy QCD processes and offer promising prospects for future precision studies in this area.
Subsequent analyses on vector quarkonia~\cite{Celiberto:2022dyf,Celiberto:2023fzz,Celiberto:2024mex,Celiberto:2024bxu}, $B_c$ mesons~\cite{Celiberto:2022keu,Celiberto:2024omj}, and $\XQq$ tetraquarks~\cite{Celiberto:2023rzw}, confirmed that this remarkable feature, known as \emph{natural stability} of the BFKL resummation~\cite{Celiberto:2022grc}, comes out as basic property connected to the inclusive emission of any given heavy-flavored particle.

\begin{figure*}[tb]
\centering

\includegraphics[width=0.45\textwidth]{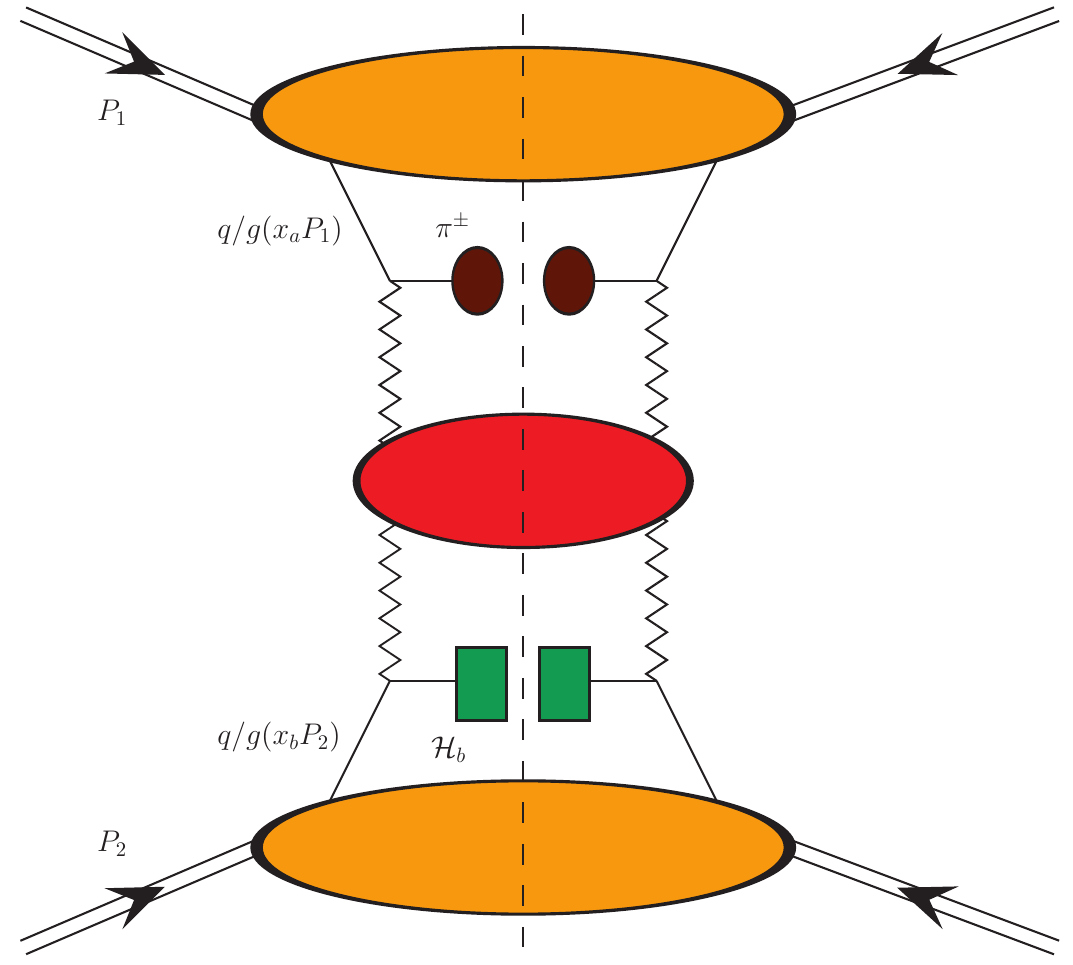}
\hspace{0.25cm}
\includegraphics[width=0.45\textwidth]{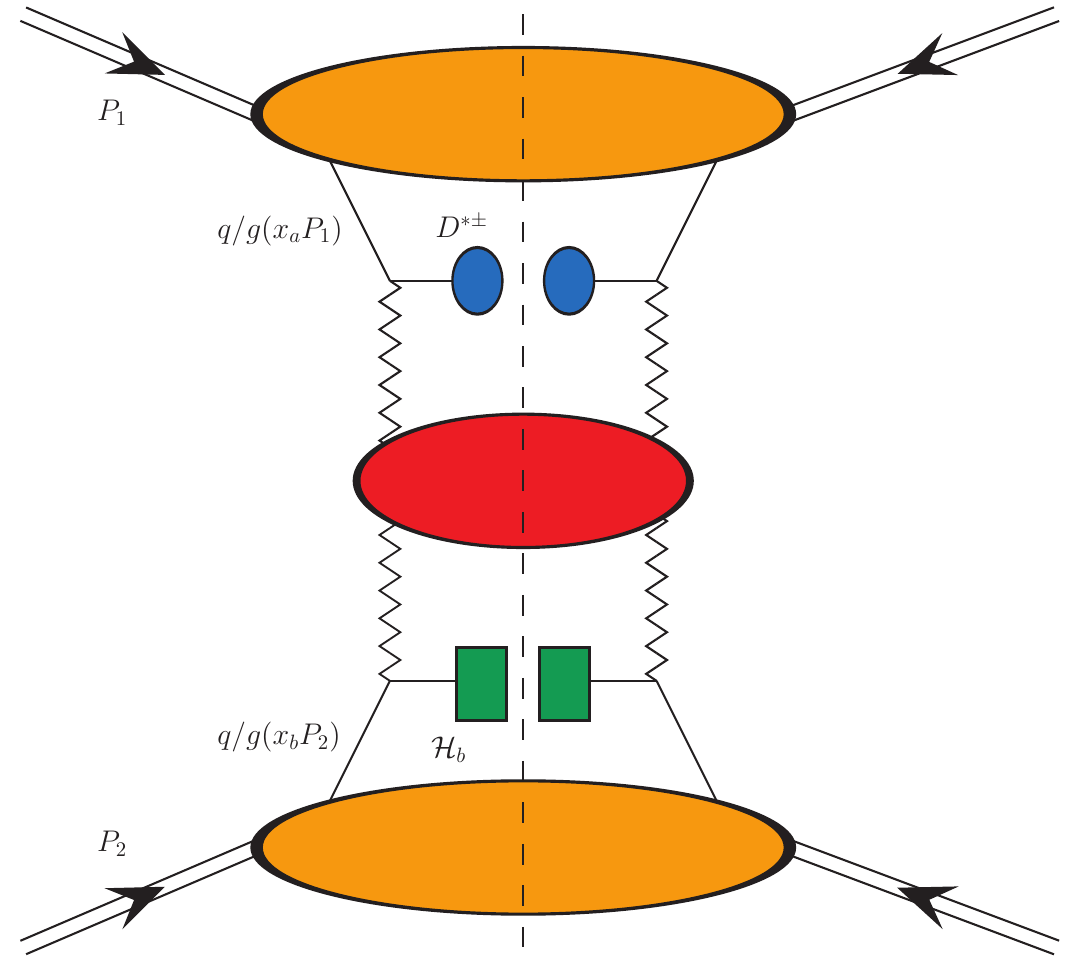}
\\ \vspace{0.25cm}
\hspace{-0.80cm}
($a$) $\pi^\pm$ $+$ $\Hb$ \hspace{5.750cm}
($b$) $D^{*\pm}$ $+$ $\Hb$

\caption{Diagrammatic representation of the hybrid factorization at work for the $\pi^\pm$ $+$ $\Hb$ ($a$, left) and $D^{*\pm}$ $+$ $\Hb$ ($b$, right) channels. The big yellow ovals denote collinear PDFs. 
Maroon (navy blue) blobs stand for $\pi^\pm$ ($D^{*\pm}$ meson collinear FFs, while green rectangles depict $\Hb$ hadron collinear FFs. 
The high-energy Green's function (red oval blob) is connected to singly off-shell coefficient functions by Reggeon zigzag lines.}
\label{fig:process}
\end{figure*}

In this review, we will provide predictions for rapidity-interval rates and angular distributions for a novel selection of forward-plus-backward two-particle semi-hard reactions. 
These processes involve final states characterized by identified hadrons only (see Fig.~\ref{fig:process}). 
Specifically, the first hadron can be a charged pion or a charged $D^*$ meson. 
The second hadron singly $b$-flavored particle, \emph{i.e.} and inclusive state consisting of the sum of fragmentation channels to noncharmed $B$ mesons and $\Lambda_b^0$ baryons.

Comparing predictions from the BFKL-driven approach with those from a high-energy fixed-order treatment will help gauge the impact of high-energy resummation on top of the DGLAP approach.
To this end, a numerical tool for calculating NLO cross sections for the inclusive production of two identified hadrons widely separated in rapidity in proton collisions would be needed and crucial for a systematic high-energy versus DGLAP analysis. 
While the LO limit for such reactions can be extracted from higher-order works~\cite{Chiappetta:1996wp,Owens:2001rr,Binoth:2002ym,Binoth:2002wa,Almeida:2009jt,Hinderer:2014qta}, it currently cannot be compared with calculations based on our hybrid factorization due to kinematic constraints. Two-particle LO computations without resummation typically result in a back-to-back final state, which is incompatible with the asymmetric windows for observed transverse momenta in our calculations.

Therefore, to assess the weight of the NLL resummation on top of the DGLAP approach, we will compare BFKL predictions with the corresponding ones obtained from a high-energy fixed-order treatment, firstly developed to address light-flavored di-jet~\cite{Celiberto:2015yba,Celiberto:2015mpa} and hadron-plus-jet~\cite{Celiberto:2020wpk} studies.
It builds upon truncating of the high-energy series up to the NLO level, so that we can reconstruct the high-energy signal of a pure NLO computation.

We will consider two different rapidity configurations.
The first one is a standard LHC tagging, where both particles ($\pi^\pm$ or $D^{*\pm}$ and the $b$-hadron) are detected by a current LHC detector, say CMS or ATLAS.
This selection provides symmetric rapidity windows for both particles, making it an ideal channel for further testing high-energy QCD dynamics, akin to previous investigations into two-particle semi-hard processes.
In the second scenario, we consider the tag of the pion or the $D$ meson in a far-forward rapidity range accessible at future \ac{FPFs}, while the bottomed hadron is simultaneously detected in the barrel of a current LHC detector. Our exploration fairly takes inspiration for a prospective FPF~$+$~ATLAS study~\cite{Celiberto:2022rfj,Celiberto:2022zdg} proposed in the context of the FPF program~\cite{Anchordoqui:2021ghd,Feng:2022inv} (see Refs.~\cite{Arakawa:2022rmp,Maciula:2022lzk,Bhattacharya:2023zei,Fieg:2023kld,Cruz-Martinez:2023sdv,Buonocore:2023kna,Wilkinson:2023vvu,Feng:2024zfe} for related work).

The simultaneous tag of a far-forward object and a central one results in an asymmetric configuration between the longitudinal fractions $x$ of the two struck partons. 
One parton has a large $x$, while the other assumes more moderate $x$ values. 
Thus, a coincidence between the FPF and a LHC detector give us a faultless chance to unravel not only the high-energy dynamics arising from to the very large rapidity intervals accessed, but also the connection between BFKL and threshold resummations. This interplay can shed light on the dynamics of parton interactions in this kinematic regime, providing valuable insights into the underlying physics.

The impact of double BFKL-plus-threshold logarithms for central Higgs-bosons inclusive rates is small at ongoing LHC energies, but it becomes quite relevant at the nominal ones of the \ac{FCC}~\cite{Mangano:2016jyj,FCC:2018byv,FCC:2018evy,FCC:2018vvp,FCC:2018bvk}.
On the other side, the high-energy resummed TM rates for the Higgs-plus-jet hadroproduction already deviate from the fixed-order background at current LHC energies~\cite{Celiberto:2020tmb}.
Thus, our two-particle observables are expected to be very sensitive to the co-action of the two resummations.

Future analyses at FPFs will be crucial for deepening our understanding of perturbative QCD and the structure of protons and nuclei in previously unexplored regimes. 
The sensitivity of FPF detectors to far-forward production of light hadrons and charmed mesons will allow us to investigate BFKL effects and gluon-recombination dynamics. 
Additionally, TeV-scale neutrino-induced DIS experiments at FPFs will serve as valuable probes of proton structure and the production mechanisms of heavy or light decaying hadrons. 
Our work on light mesons at the FPF via hybrid factorization can provide a common framework for describing the production and decays of these particles.

QCD studies are fundamental components of the multi-frontier research program at FPFs. Searches for long-lived particles, indirect detection of dark matter, sterile neutrinos, as well as investigations into the muon puzzle, lepton universality, and the connection between high-energy particle physics and modern astroparticle physics, all depend on a deep understanding of the Standard Model. Progress towards precision QCD in the kinematic sectors reachable at FPFs will be essential for driving scientific interest towards new and compelling directions.

This review is organized as follows: highlights on the $\NLLp$ hybrid factorization for our reference processes are given in Section~\ref{sec:theory}; the phenomenological analysis is discussed in Section~\ref{sec:phenomenology}; conclusions and outlook are drawn in Section~\ref{sec:conclusions}.

%-----------------------------------------
\section{Hybrid factorization at work}
\label{sec:theory}
%-----------------------------------------

In this Section we presents basic features of the $\NLLp$ hybrid factorization well-adapted to the description of our reactions. After a brief introduction of the process kinematics~(Section~\ref{ssec:process}), we give details of the NLL resummed cross section~(Section~\ref{ssec:sigma}). Choices for collinear PDFs and FFs are explained later~(Section~\ref{ssec:PDFs_FFs}).

%-----------------------------------------
\subsection{Kinematics of the process}
\label{ssec:process}
%-----------------------------------------

We consider the two following processes (Fig.~\ref{fig:process})
\begin{equation}
\begin{split}
\label{process}
 p(P_1) + p(P_2) &\rightarrow {\pi^\pm}(q_\pi, y_\pi, \phi_\pi) + {\cal X} + \Hb(q_b, y_b, \phi_b) \; ,
\\[0.15cm]
 p(P_1) + p(P_2) &\rightarrow {D^{*\pm}}(q_D, y_D, \phi_D) + {\cal X} + \Hb(q_b, y_b, \phi_b) \; .
\end{split}
\end{equation}
The upper equation of~\eqref{process} says that a $\pi^\pm$ meson with mass $M_\pi = 139.57$~MeV, four-momentum $q_\pi$, rapidity $y_\pi$, and azimuthal angle $\phi_\pi$ is emitted in association with a singly bottomed hadron $\Hb$ with four-momentum $q_b$, rapidity $y_b$, and azimuthal angle $\phi_b$.
In the lower equation the pion is replaced by a $D^{*\pm}$ meson with mass $M_D = 1.968$~GeV, four-momentum $q_D$, rapidity $y_D$, and azimuthal angle $\phi_D$.
The $\Hb$ particle is given by an inclusive sum of all noncharmed $B$ mesons and $\Lambda_b$ baryons.

Initial-state partons posses four-momentum $x_a P_1$ and $x_b P_2$, where $P_{1,2}$ are the momenta of the incoming protons.
The $\cal X$ system in Eq.~\eqref{process} depicts all the final-state inclusively irradiated gluons.
One can take $P_{1,2}$ as Sudakov vectors satisfying $P_1^2 = P_2^2 = 0$ and $ 2 P_1 \cdot P_2 = s$, to decompose the final-state four momenta as
\begin{equation}\label{sudakov}
q_{{\cal M},b} = x_{{\cal M},b} P_{1,2} - \frac{\vec q_{\perp_{{\cal M},b}}^{\,2}}{x_{{\cal M},b}}\frac{P_{2,1}}{s} + q_{\perp_{{\cal M},b}} \,,
\qquad
q_{\perp_{{\cal M},b}}^2 = -\vec q_{T_{{\cal M},b}}^{\,2} \;,
\end{equation}
where the ${\cal M}$ subscript inclusively refer to $\pi$ or $D$.
The outgoing-object longitudinal momentum fractions, $x_{{\cal M},b}$, can be calculated by inverting the relation
\begin{equation}\label{yMH}
y_{{\cal M},b} = 
\pm \ln \left( \frac{x_{{\cal M},b}}{|\vec q_{T_{{\cal M},b}}|} \sqrt{s} \right)
\,,
\end{equation}
so that
\begin{equation}\label{dyMH}
 \drv y_{{\cal M},b} = \pm \frac{\drv x_{{\cal M},b}}{x_{{\cal M},b}}
\;.
\end{equation}
The semi-hard nature of our reactions
follows $(i)$ from the size of final-state transverse momenta, which are taken to respect the hierarchy $\Lambda_{\rm QCD} \ll |\vec q_{T_{{\cal M},b}}| \ll \sqrt{s}$, and $(ii)$ from imposing a large rapidity distance $\DY = y_{\cal M} - y_b$ between the ${\cal M}$ meson and the $b$-hadron. 
Furthermore, to ensure the validity of a VFNS description for heavy-hadron production~\cite{Mele:1990cw,Cacciari:1993mq,Buza:1996wv}, $|\vec q_{T_{{\cal M},b}}|$ ranges must remain sufficiently above the thresholds for DGLAP evolution determined by the charm and bottom masses.

%-----------------------------------------
\subsection{NLO cross section resummed at NLL and beyond}
\label{ssec:sigma}
%-----------------------------------------

Following a pure QCD collinear vision for the LO cross section of our processes (Eq.~\eqref{process}), one would take a one-dimensional convolution between the partonic hard subprocess, incoming proton PDFs and tagged hadron FFs
\begin{equation}
\label{sigma_collinear_LO}
\begin{split}
\hspace{-0.05cm}
\frac{\drv \sigma^{[p + p \,\rightarrow\, {{\cal M}} + {\Hb} + {\cal X}]}_{\rm LO,\,collinear}}{\drv x_{\cal M}\drv x_{b}\drv ^2\vec q_{T_{\cal M}}\drv ^2\vec q_{T_{b}}}
&\,=\, \sum_{i,j=q,{\bar q},g}\int_0^1 \drv x_a \int_0^1 \drv x_b\ f_i\left(x_a,\mu_F\right) f_j\left(x_b,\mu_F\right)
\\
&\,\times\, \int_{x_{\cal M}}^1\frac{\drv \zeta_1}{\zeta_1}\int_{x_{b}}^1\frac{\drv \zeta_2}{\zeta_2}D^{{\cal M}}_{i}\left(\frac{x_{\cal M}}{\zeta_1},\mu_F\right)D^{b}_{j}\left(\frac{x_{b}}{\zeta_2},\mu_F\right)
\frac{\drv {\hat\sigma}_{i,j}\left(\hat s,\mu_F,\mu_R\right)}
{\drv x_{\cal M}\drv x_{b}\drv ^2\vec q_{T_{\cal M}}\drv ^2\vec q_{T_{b}}}\;.
\end{split}
\end{equation}
Here the $i,j$ indices run over all the parton species except for the top quark which does not hadronize, with $f_{i,j}\left(x_{1,2}, \mu_F \right)$ being proton PDFs and $D^{{\cal M},b}_{i,j}\left(x_{{\cal M},b}/\zeta_{1,2}, \mu_F \right)$ standing for ${{\cal M}}$ meson and $b$-hadron FFs.
Then, $x_{1,2}$ are longitudinal-momentum fractions for the partons initiation the hard scattering and $\zeta_{1,2}$ the ones for partons fragmenting to detected hadrons. 
The hard factor $\hat \sigma_{i,j} \left( \hat s,\mu_F,\mu_R \right)$ depends on the squared center-of-mass energy of the partonic collision, $\hat s \equiv x_a x_b s$, and on factorization ($\mu_F$) and renormalization ($\mu_R$) scales.

\emph{Vice versa}, to construct differential observables in our $\NLLp$ hybrid formalism one first performs the high-energy resummation of logarithms connected to transverse-momentum exchanges in the $\mbox{\emph{t}-channel}$. 
Subsequently, one adds collinear ingredients, namely PDFs and FFs.
The $\NLLp$ differential cross section can be rewritten as a Fourier sum of angular coefficients
\begin{equation}
 \label{dsigma_Fourier}    
 %\hspace{-0.19cm}
 \frac{\drv \sigma_{\NLLp}^{[p + p \,\rightarrow\, {{\cal M}} + {\cal X} + {\Hb}]}}{\drv y_{\cal M} \, \drv y_{b} \, \drv|\vec q_{T_{\cal M}}| \, \drv|\vec q_{T_{b}}| \, \drv\phi_{{{\cal M}}} \, \drv\phi_{b}} = 
 \frac{1}{(2{\pi})^2} \left[{\cal C}_0 + 2 \sum_{n=1}^\infty \cos (n \varphi)\,
 {\cal C}_n \right]\, ,
\end{equation}
where $\varphi = \phi_{\cal M} - \phi_{b} - \pi$ is the distance between the azimuthal angles of the light and the heavy hadron.
The angular coefficients ${\cal C}_n$ are calculated within the BFKL framework and they embody the resummation of energy logarithms. A NLL-consistent formula obtained in the \ac{MSb} renormalization scheme~\cite{PhysRevD.18.3998} is cast as follows (for details on the derivation see, \emph{e.g.}, Ref.~\cite{Caporale:2012ih})
\begin{equation}
\label{Cn_NLLp_MSb}%\nonumber
\begin{split}
 \CnNLLp &\,=\, \int_0^{2\pi} \drv \phi_{\cal M} \int_0^{2\pi} \drv \phi_{b}\,
 \cos (n \varphi) \,
 \frac{\drv \sigma^{[p + p \,\rightarrow\, {{\cal M}} + {\cal X} + \Hb]}_\NLLp}{\drv y_{\cal M} \, \drv y_{b} \, \drv |\vec q_{T_{\cal M}}| \, \drv |\vec q_{T_{b}}| \, \drv \phi_{\cal M} \, \drv \phi_{b}}\;
\\
 &\,=\, \frac{e^{\DY}}{s} %\frac{x_{\cal M} x_{b}}{p_{\cal M} p_{b}}
 \int_{-\infty}^{+\infty} \drv \nu \, e^{{\DY} \bar \alpha_s(\mu_R) \chi^{\rm NLL}(n,\nu)}
 \alpha_s^2(\mu_R)
%\[
% = \frac{e^{\DY}}{s} %\frac{x_{\cal M} x_{b}}{p_{\cal M} p_{b}}
% \int_{-\infty}^{+\infty} \drv \nu \, e^{{\DY} \bar \alpha_s(\mu_R)\left\{\chi(n,\nu)+\bar\alpha_s(\mu_R)
% \left[\bar\chi(n,\nu)+\frac{\beta_0}{8 N_c}\chi(n,\nu)\left[-\chi(n,\nu)+\frac{10}{3}+4\ln\left(\frac{\mu_R}{\sqrt{|\vec q_{T_{\cal M}}| |\vec q_{T_{b}}|}}\right)\right]\right]\right\}}
%\]
%\[
% \times \, \alpha_s^2(\mu_R) c^{{\cal M}}(n,\nu,|\vec q_{T_{\cal M}}|, x_{\cal M})[c^{b}(n,\nu,|\vec q_{T_{b}}|,x_{b})]^*\,
%\]
\\
 &\,\times\, %\alpha_s^2(\mu_R) \, 
 \left[
 c_{\cal M}^{\rm NLO}(n,\nu,|\vec q_{T_{\cal M}}|, x_{\cal M}) \,
 [c_{b}^{\rm NLO}(n,\nu,|\vec q_{T_{b}}|,x_{b})]^*\,
% \times \, 
 + \bar \alpha_s^2(\mu_R) 
 \, \DY
 \frac{\beta_0}{4 N_c}\chi(n,\nu)\,\F(\nu)
 \right] \;,
%\begin{equation}
%\label{Cn_NLLp_MSb}%\nonumber
% \times \, \left\{1
% +\alpha_s(\mu_R)\left[\frac{\hat c^{{\cal M}}(n,\nu,|\vec q_{T_{\cal M}}|,x_{\cal M})}{c^{{\cal M}}(n,\nu,|\vec q_{T_{\cal M}}|,x_{\cal M})}
% +\left[\frac{\hat c^{b}(n,\nu,|\vec q_{T_{b}}|, x_{b})}{c^{b}(n,\nu,|\vec q_{T_{b}}|,x_{b})}\right]^*
% +  \bar \alpha_s(\mu_R) %\alpha_s^2(\mu_R) 
% \, \DY
% \frac{\beta_0}{4 {\cal M}}\chi(n,\nu)f(\nu)
% \right]
% \right\} \;,
%\end{equation}
\end{split}
\end{equation}
with $\bar \alpha_s(\mu_R) \equiv \alpha_s(\mu_R) N_c/\pi$, $N_c$ the number of colors and $\beta_0$ the QCD $\beta$-function leading coefficient.
The $\chi^{\rm NLL}\left(n,\nu\right)$ stands for the NLL high energy kernel:
\begin{equation}
\chi^{\rm NLL}(n,\nu) = \chi(n,\nu) +\bar\alpha_s(\mu_R) \left[\bar\chi(n,\nu)+\frac{\beta_0}{8 N_c}\chi(n,\nu)\left[-\chi(n,\nu)+\frac{10}{3}+2\ln \frac{\mu_R^2}{|\vec q_{T_{\cal M}}| |\vec q_{T_{b}}|} \right]\right] \;,
\label{chi_NLO}
\end{equation}
where
\begin{equation}
%\hspace{-0.45cm}
\chi\left(n,\nu\right)=2\left\{\psi\left(1\right)-{\rm Re} \left[\psi\left( i\nu+\frac{n+1}{2} \right)\right] \right\}
\label{chi}
\end{equation}
is its eigenvalue at LL and $\psi(z) = \Gamma^\prime(z)/\Gamma(z)$. 
The $\hat\chi(n,\nu)$ NLO term was calculated in Ref.~\cite{Kotikov:2000pm,Kotikov:2002ab} and can be found in the Appendix~\hyperlink{app:NLL_kernel}{A}.
The $c_h^{\rm NLO}(n,\nu,|\vec q_T|, x)$ emission function describe the forward inclusive emission of a given hadron, labeled as $h$. 
It was obtained in Ref.~\cite{Ivanov:2012iv} in the light-quark cases and embodies collinear inputs.
It also can be employed for heavy hadrons within a VFNS scheme, provided that observed transverse momenta $|\vec q_{T_{b}}|$ are sufficiently higher than the charm (for a $D$ meson) or bottom (for a $b$~hadron) masses.
One has
\begin{equation}
\label{HIF}
c_h^{\rm NLO}(n,\nu,|\vec q_T|, x) =
c_h(n,\nu,|\vec q_T|, x) +
\alpha_s(\mu_R) \, \hat c_h(n,\nu,|\vec q_T|, x) \; ,
\end{equation}
with
\begin{equation}
\label{LOHIF}
\hspace{-0.25cm}
c_h(n,\nu,|\vec q_T|, x) 
= 2 \sqrt{\frac{C_F}{C_A}}
|\vec q_T|^{2i\nu-1} \int_{x}^1\frac{\drv \zeta}{\zeta}
\left( \frac{\zeta}{x} \right)
^{2 i\nu-1} 
% \times 
 \left[\frac{C_A}{C_F}f_g(\zeta)D_g^{h}\left(\frac{x}{\zeta}\right)
 +\sum_{i=q,\bar q}f_i(\zeta)D_i^{h}\left(\frac{x}{\zeta}\right)\right] %\;.
\end{equation}
%\end{widetext}
its LO limit and $\hat c_h(n,\nu,|\vec q_T|, x)$ its NLO correction (see Appendix~\hyperlink{app:NLO_IF}{B}).
The $\F(\nu)$ function in Eq.~\eqref{Cn_NLLp_MSb} embodies the logarithmic derivative of the two LO emission functions
%\begin{widetext}
\begin{equation}
 \F(\nu) = \frac{i}{2} \, \frac{\drv}{\drv \nu} \ln\left(\frac{c_{\cal M}(n,\nu,|\vec q_{T_{b}}|, x_{\cal M})}{[c_{b}(n,\nu,|\vec q_{T_{b}}|, x_{b})]^*}\right) + \ln\left(|\vec q_{T_{\cal M}}| |\vec q_{T_{b}}|\right) \;.
\label{fnu}
\end{equation}
From Eqs.~(\ref{Cn_NLLp_MSb})-(\ref{Cn_LL_MSb}), we gather the implementation of our hybrid factorization approach. 
The cross section is expressed as a factorized formula reminiscent of the BFKL formalism. 
In this formulation, the Green's function read as a high-energy convolution between the emission functions of the two tagged hadrons. 
These functions are further expressed as collinear convolutions between PDFs and FFs, along with the hard-scattering term.
The `$+$' label indicates that our representation for angular coefficients in Eq.~\eqref{Cn_NLLp_MSb} encode terms beyond the NLL accuracy generated by the cross product of the NLO emission functions, $\hat c_{\cal M}(n,\nu,|\vec q_{T_{\cal M}}|, x_{\cal M}) \, [\hat c_{b}(n,\nu,|\vec q_{T_{b}}|,x_{b})]^*$.

Expanding and truncating to the ${\cal O}(\alpha_s^3)$ order the $\NLLp$ angular coefficients in Eq.~(\ref{Cn_NLLp_MSb}), one gets the high-energy fixed-order ($\HENLOp$) counterpart of the BFKL-resummed cross section~\cite{Celiberto:2015yba,Celiberto:2015mpa,Celiberto:2020wpk,Celiberto:2020rxb,Celiberto:2021dzy}).
It is sensitive to the leading-power asymptotic dynamics of a pure NLO DGLAP computation and discards factors suppressed by inverse powers of $\hat s$.
Our $\HENLOp$ expression for angular coefficients in the $\MSb$ scheme~\cite{PhysRevD.18.3998} reads
\begin{equation}
\label{Cn_HENLOp_MSb}%\nonumber
 \CnHENLOp = \frac{e^{\Delta Y}}{s}
 \int_{-\infty}^{+\infty} \drv \nu \, \alpha_s^2(\mu_R) \,
 c_{\cal M}^{\rm NLO}(n,\nu,|\vec q_{T_{\cal M}}|,x_{\cal M}) \,
 [c_{b}^{\rm NLO}(n,\nu,|\vec q_{T_{b}}|,x_{b})]^*
 \left[ 1 + \bar \alpha_s(\mu_R) \DY \chi(n,\nu) \right] \,.
\end{equation}
In our approach, an expansion up to contributions proportional to $\alpha_s(\mu_R)$ takes over the BFKL exponentiated kernel. Analogous to Eq.~\eqref{Cn_NLLp_MSb}, our fixed-order formula at high energies is denoted as $\HENLOp$, meaning contributions the beyond NLL accuracy arise from the cross product of NLO emission functions.

Then, we can obtain a genuine $\LL$ formula by simply discarding all the NLO contributions, thus having
\begin{equation}
\label{Cn_LL_MSb}%\nonumber
  \CnLL = \frac{e^{\DY}}{s} %\frac{x_{\cal M} x_{b}}{p_{\cal M} p_{b}}
 \int_{-\infty}^{+\infty} \drv \nu \, e^{{\DY} \bar \alpha_s(\mu_R)\chi(n,\nu)} \, \alpha_s^2(\mu_R) \, c(n,\nu,|\vec q_{T_{\cal M}}|, x_{\cal M}) \,
 [c(n,\nu,|\vec q_{T_{b}}|,x_{b})]^* \,.
\end{equation}

In our phenomenological analysis (see Section~\ref{sec:phenomenology}) we compare observables built in terms of $\NLLp$, $\HENLOp$, and $\LL$ angular coefficients. We fix $\mu_{R,F}$ scales at the \emph{natural} energies given by kinematics, thus having $\mu_R = \mu_F = \mu_N \equiv m_{\perp_{\cal M}} + m_{\perp_{b}}$, with $m_{h \perp} = \sqrt{m_h^2 + |\vec q_{T_{h}}|^2}$ being the transverse mass of the specific hadron $h$. 
To assess the impact of MHOUs, scales will be varied from $1/2$ to $2$ times their natural values, as specified by the $C_\mu$ parameter (see Section~\ref{ssec:uncertainty}).

We employ a two-loop QCD-coupling setup with $\alpha_s\left(M_Z\right)=0.11707$ and five quark flavors active. 
In the $\MSb$ renormalization scheme~\cite{PhysRevD.18.3998}, we write
\begin{equation}
\label{as_MSb}
 \as(\mu_R) \equiv \as^{\MSb}(\mu_R) = \frac{\pi}{\beta_0 \, \lambda_R} \left( 4 - \frac{\beta_1}{\beta_0^2} \frac{\ln \lambda_R}{\lambda_R} \right) %\;,
\end{equation}
with
\begin{equation}
\label{as_parameters}
 \lambda_R(\mu_R) = \ln \frac{\mu_R^2}{\LQCD^2} \;,
 \qquad
 \beta_0 = 11 - \frac{2}{3} n_f \;, 
 \qquad
 \beta_1 = 102 - \frac{38}{3} n_f \;.
\end{equation}

We emphasize that in our approach, energy scales are inherently tied to the transverse masses of observed particles. Consequently, they consistently fall within the perturbative regime, obviating the need for any infrared enhancement of the running coupling (see, for instance, Ref.~\cite{Webber:1998um}). Moreover, the utilization of large-scale values shields us from a regime where the significance of the \emph{diffusion pattern} (see, for instance, Refs.~\cite{Bartels:1993du,Caporale:2013bva,Ross:2016zwl}) becomes pronounced.

%-----------------------------------------
\subsection{Choice of collinear PDFs and FFs}
\label{ssec:PDFs_FFs}
%-----------------------------------------

As we already mentioned, due to the moderate parton $x$~values, we build our framework upon standard collinear inputs, at NLO while the high-energy resummation is performed via BFKL at NLL.
As for proton PDFs, make us of the novel {\tt NNPDF4.0} NLO determination~\cite{NNPDF:2021uiq,NNPDF:2021njg}.

When considering charged pions, there is a broad array of NLO FFs available for potential use.
The {\tt NNFF1.0} FFs~\cite{Bertone:2017tyb,Bertone:2018ecm}, derived from SIA data using a neural-network approach, feature NLO gluon FFs.
On the other hand, {\tt DEHSS14} sets~\cite{deFlorian:2014xna} were obtained from a combination of SIA, \ac{SIDIS}, and proton-proton collision data. They assume a partial $SU(2)$ isospin symmetry, resulting in specific relations among the different flavor components.
Meanwhile, {\tt JAM20} determinations~\cite{Moffat:2021dji} incorporate datasets from both SIA and SIDIS and are determined concurrently with collinear PDFs in DIS and fixed-target Drell-Yan measurements. These FFs adhere to a full $SU(2)$ isospin symmetry.

Recently, {\tt MAPFF1.0} functions~\cite{Khalek:2021gxf,Khalek:2022vgy} have been derived from SIA and SIDIS data using neural-network techniques. Notably, these FFs allow for a deviation from isospin symmetry, with separate parametrizations for $D_u^{\pi^+}$ and $D_{\bar d}^{\pi^+}$, which varies with the momentum fraction $z$. Gluon FFs are generated at NLO, although the data are collected at lower energies, where the gluon distribution has a more pronounced impact.
Moreover, the methodology employed for extracting {\tt MAPFF1.0} FFs has been extended to study the fragmentation of the $\Xi^-/{\bar \Xi}^+$ octet baryon~\cite{Soleymaninia:2022qjf} and to establish a new FF set for describing unidentified charged light hadrons, both from SIA and SIDIS data~\cite{Soleymaninia:2022alt}.
In our phenomenological analysis, we make use of two pion NLO FF determinations: {\tt NNFF1.0} and {\tt MAPFF1.0}.

As regards heavy hadrons, we employ the {\tt KKKS08} NLO set~\cite{Kniehl:2004fy,Kniehl:2005de,Kniehl:2006mw,Kneesch:2007ey} to describe parton fragmentation into $\Lambda_c$ baryons.
These FFs were extracted from OPAL and Belle data for SIA and mainly rely on a Bowler-like description~\cite{Bowler:1981sb} for charm and bottom flavors.
We depict emissions of $b$~flavored hadrons in terms of the {\tt KKSS07} parametrization~\cite{Kniehl:2008zza} based on data of the inclusive $B$-meson production in SIA events at CERN LEP1 and SLAC SLC and portrayed by a simple, three-parameter power-like function~\cite{Kartvelishvili:1985ac} for heavy-quark species.
The {\tt KKSS19} and {\tt KKSS07} determinations use the VFNS. 
We remark that the employment of given VFNS PDFs or FFs is admitted in our approach, provided that typical energy scales are much larger than thresholds for the DGLAP evolution of charm and bottom quarks. As highlighted in Section~\ref{ssec:final_state}, this requirement is always fulfilled.\footnote{For further studies on $D$-meson, $\Lambda_c$-baryon and $b$-hadron VFNS fragmentation, see Refs.~\cite{Corcella:2007tg,Anderle:2017cgl,Salajegheh:2019nea,Salajegheh:2019srg,Soleymaninia:2017xhc},~\cite{Binnewies:1998vm,Kniehl:2007erq,Kniehl:2011bk,Kniehl:2012mn,Kramer:2018vde,Kramer:2018rgb,Salajegheh:2019ach,Kniehl:2021qep}, and~\cite{Kniehl:2005de,Kniehl:2020szu,Delpasand:2020vlb}, respectively.}

We note that {\tt KKSS08} and {\tt KKSS07} FF sets do not carry any quantitative information regarding the extraction uncertainty. Future investigations incorporating potential new parametrizations of $\Lambda_c$ and $\Hb$ fragmentation functions, including uncertainties, are essential to supplement our analysis of systematic errors in high-energy distributions.

%-----------------------------------------
\section{Heavy hadrons with {\Jethad}}
\label{sec:phenomenology}
%-----------------------------------------

All predictions presented in this review were generated using \textsc{Jethad}, a hybrid code that consistently integrates both \textsc{Python}- and \textsc{Fortran}-based modules. 
{\Jethad} is specifically designed for computing, managing, and processing physical distributions defined within various formalisms~\cite{Celiberto:2020wpk,Celiberto:2022rfj,Celiberto:2023fzz,Celiberto:2024mrq}. 
Numeric calculations of differential distributions were primarily performed using \textsc{Fortran 2008} modular routines within {\Jethad}, while the built-in \textsc{Python 3.0} analyzer was utilized for final data analysis and interpretation.
All computation of observables are conducted within the $\MSb$ scheme~\cite{PhysRevD.18.3998}.

Section~\ref{ssec:jethad} briefly introduces core elements of the {\tt v0.5.2} version of {\Jethad}.
Our strategy to gauge systematic uncertainties is discussed in Section~\ref{ssec:uncertainty}. 
Final-state kinematic ranges can be found in Section~\ref{ssec:final_state}.
Predictions rapidity-interval rates and angular multiplicities are discussed in Sections~\ref{ssec:Y_rates} and~\ref{ssec:phi_multiplicities}, respectively.

%-----------------------------------------
\subsection{Highlights of {\Jethad} {\tt v0.5.2}}
\label{ssec:jethad}
%-----------------------------------------

The inception of the {\Jethad} project dates back to late 2017, driven by the necessity for precise predictions of semi-hard hadron~\cite{Celiberto:2016hae,Celiberto:2017ptm} and jet~\cite{Celiberto:2015yba,Celiberto:2016ygs,Bolognino:2018oth} sensitive final states at the LHC.
Phenomenological analyses of such reactions, proposed as probe channels for high-energy resummation in QCD, required the development of a reference numerical framework dedicated to computing and analyzing high-energy related distributions.

The initial named version, {\Jethad} {\tt v0.2.7}, provided us with a first, quantitative BFKL-versus-DGLAP examination within the context of semi-inclusive hadron-plus-jet emissions at the LHC~\cite{Celiberto:2020wpk}. Subsequent iterations introduced new functionalities, such as selecting forward heavy-quark pair observables ({\tt v0.3.0}~\cite{Bolognino:2019yls}), enabling studies on Higgs emissions and transverse-momentum distributions ({\tt v0.4.2}~\cite{Celiberto:2020tmb}), and integrating the \textsc{Python} analyzer with the \textsc{Fortran} core supermodule ({\tt v0.4.3}~\cite{Bolognino:2021mrc}).

Advancements continued with the ability to analyze heavy-flavored hadrons via VFNS FFs at NLO ({\tt v0.4.4}~\cite{Celiberto:2021dzy}). The \deffont{\rlap{D}Unamis} (\textsc{DYnamis}) work package, dedicated to the forward DY dilepton reaction~\cite{Celiberto:2018muu}, became part of {\Jethad} in {\tt v0.4.5}. Integration with the \textsc{LExA} modular code enabled exploration of proton content at low-$x$ through small-$x$ TMD densities in {\tt v0.4.6}~\cite{Bolognino:2021niq}.

Version {\tt v0.4.7}~\cite{Celiberto:2022dyf} introduced quarkonium-sensitive reactions from NRQCD leading-twist fragmentation. 
Novel features in {\tt v0.5.0}~\cite{Celiberto:2023fzz} and {\tt v0.5.1}~\cite{Celiberto:2024omj} encompass an enhanced system for MHOU-related studies, an expanded list of observables with a focus on singly- and doubly-differential transverse-momentum production rates~\cite{Celiberto:2022gji,Celiberto:2022kxx,Celiberto:2024omj}, and support for \emph{matching} procedures with collinear factorization~\cite{Celiberto:2023rtu,Celiberto:2023uuk,Celiberto:2023eba,Celiberto:2023nym,Celiberto:2023rqp}.
The most significant update of~{\tt v0.5.2} is {\symJethad}, a \textsc{Mathematica} plugin devoted to symbolical calculations for high-energy QCD and the proton structure.

From the fundamental core to service modules and routines, {\Jethad} has been designed to dynamically achieve high levels of computational performance. The multidimensional integrators within {\Jethad} support extensive parallel computing to actively choose the most suitable integration algorithm based on the shape of the integrand.

Any process implemented in {\Jethad} can be dynamically selected through an intuitive, \emph{structure} based smart-management interface. Physical final-state particles are represented by \emph{object} prototypes within this interface, where particle objects encapsulate all pertinent information about their physical counterparts, ranging from mass and charge to kinematic ranges and rapidity tags. These particle objects are initially loaded from a master database using a dedicated \emph{particle generation} routine, and custom particle generation is also supported. Then, these objects are \emph{cloned} into a final-state vector and \emph{injected} from the integrand routine to the corresponding, process-specific module by a dedicated \emph{controller}.

The flexibility in generating the physical final states is accompanied by a range of options for selecting the initial state. A unique \emph{particle-ascendancy} structure attribute enables {\Jethad} to rapidly learn whether an object is hadroproduced, electroproduced, photoproduced, etc. This dynamic feature ensures that only relevant modules are initialized, optimizing computing-time efficiency.

{\Jethad} is structured as an \emph{object-based} interface that is entirely independent of the specific reaction under investigation. While originally inspired by high-energy QCD and TMD factorization phenomenology, the code's design allows for easy encoding of different approaches by simply implementing novel, dedicated (super)modules. These can be straightforwardly linked to the core structure of the code by means of a natively-equipped \emph{point-to-routine} system, making {\Jethad} a versatile, particle-physics oriented environment.

With the aim of providing the Scientific Community with a standard computation technology tailored for the management of diverse processes (described by distinct formalisms), we envision releasing the first public version of {\Jethad} in the medium-term future.

%-----------------------------------------
\subsection{Uncertainty estimation}
\label{ssec:uncertainty}
%-----------------------------------------

A commonly adopted approach to assess the impact of MHOUs involves examining the sensitivity of our observables to variations in the renormalization scale and the factorization scale around their natural values.
It is widely acknowledged that MHOUs significantly contribute to the overall uncertainty~\cite{Celiberto:2022rfj}. To gauge their influence, we simultaneously vary $\mu_R$ and $\mu_F$ around $\mu_N/2$ and $2 \mu_N$, with the $C_{\mu}$ parameter in the figures of Section~\ref{sec:phenomenology} denoted as $C_\mu \equiv \mu_{F}/\mu_N = \mu_{R}/\mu_N$.

Another potential source of uncertainty arises from proton PDFs. Recent analyses of high-energy production rates suggest that selecting different PDF parametrizations and members within the same set has minimal effect~\cite{Bolognino:2018oth,Celiberto:2020wpk,Celiberto:2021fdp,Celiberto:2022rfj}. Hence, our observables will be computed using only the central member of the {\tt NNPDF4.0} parametrization.

Additional uncertainties may stem from a \emph{collinear improvement} of the NLO kernel, which entails incorporating renormalization-group (RG) terms to align the BFKL equation with the DGLAP one in the collinear limit, or from changes in the renormalization scheme~\cite{Salam:1998tj,Ciafaloni:2003rd,Ciafaloni:2003ek,Ciafaloni:2000cb,Ciafaloni:1999yw,Ciafaloni:1998iv,SabioVera:2005tiv}. The impact of collinear-improvement techniques on semi-hard rapidity-differential rates is observed to be encapsulated within the error bands generated by MHOUs~\cite{Celiberto:2022rfj}.

Furthermore, the $\MSb$~\cite{PhysRevD.18.3998} to \ac{MOM}~\cite{Barbieri:1979be,PhysRevLett.42.1435} renormalization-scheme transition was assessed in Ref.~\cite{Celiberto:2022rfj}, resulting in systematically higher MOM results for rapidity distributions. However, these outcomes remain within the MHOUs bands. Notably, a proper MOM analysis should be grounded on MOM-evolved PDFs and FFs, which are presently unavailable.

To establish uncertainty bands for our distributions, we combine MHOUs with the numerical errors arising from multidimensional integration (see Section~\ref{ssec:final_state}). The latter is consistently kept below $1\%$ owing to the integrators employed in {\Jethad}.

%-----------------------------------------
\subsection{Final-state kinematic ranges}
\label{ssec:final_state}
%-----------------------------------------

\begin{figure*}[tb]
\centering

\includegraphics[width=0.475\textwidth]{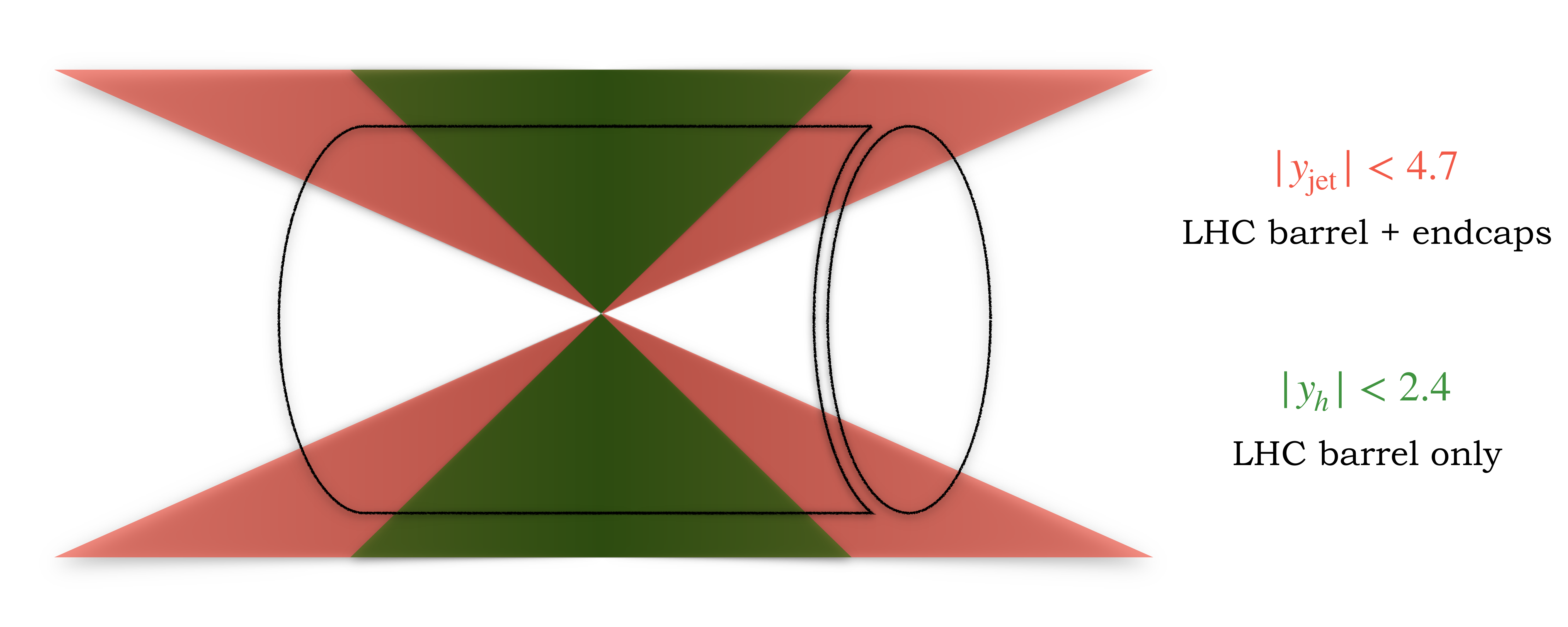}
\hspace{0.25cm}
\includegraphics[width=0.485\textwidth]{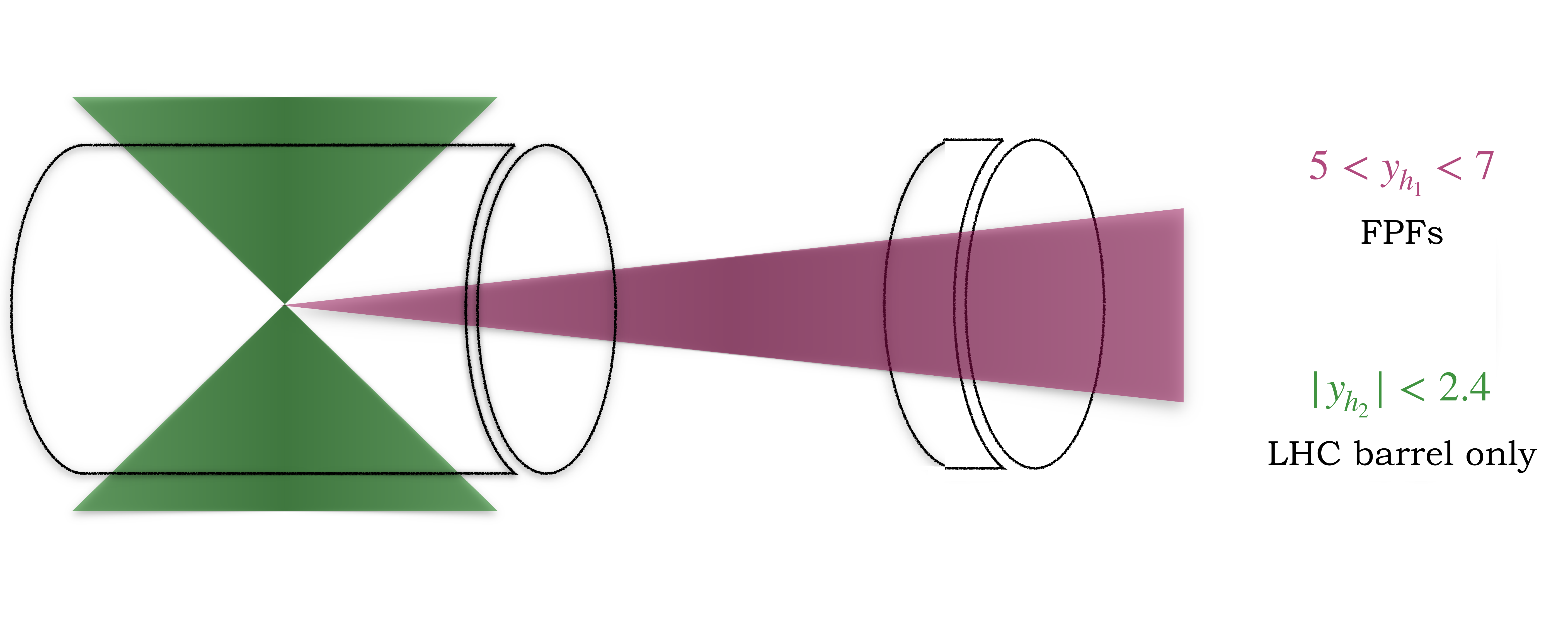}
\\ \vspace{0.25cm}
\hspace{-0.50cm}
($a$) LHC hadron-jet tagging \hspace{3.50cm}
($b$) FPF + LHC coincidence

\caption{Left: A typical forward-backward hadron-jet tag in standard LHC rapidity ranges.
Right: concurrent detection of a far-forward hadron at FPFs~\cite{Anchordoqui:2021ghd,Feng:2022inv} and a central one at an LHC experiment thanks a narrow timing coincidence.}
\label{fig:detectors}
\end{figure*}

By making use of formul{\ae} of Eqs.~(\ref{Cn_NLLp_MSb}), (\ref{Cn_HENLOp_MSb}), and~(\ref{Cn_LL_MSb}), we construct physical observables as functions of angular coefficients \emph{integrated} over the final-state phase-space variables, while the rapidity separation $\DY = y_{\cal M} - y_{b}$ between the two observed particles is kept fixed. 
Thus, we have
\begin{equation}
 \label{DY_distribution}
 C_n (\DY, s) =
 \int_{q_{T_{\cal M}}^{\rm min}}^{q_{T_{\cal M}}^{\rm max}} \drv |\vec q_{T_{\cal M}}|
 \int_{q_{T_{b}}^{\rm min}}^{q_{T_{b}}^{\rm max}} \drv |\vec q_{T_{b}}|
 \int_{y_{\cal M}^{\rm min}}^{y_{\cal M}^{\rm max}} \drv y_{\cal M}
 \int_{y_{b}^{\rm min}}^{y_{b}^{\rm max}} \drv y_{b}
 \, \,
 \delta (\DY - (y_{\cal M} - y_{b}))
 \, \,
 {\cal C}_n%\left(|\vec q_{T_{\cal M}}|, |\vec q_{T_{b}}|, y_{\cal M}, y_{b} \right)
 \, .
\end{equation}
The $C_n$ terms inclusively represents all the $\NLLp$, $\HENLOp$, and $\LL$ integrated angular coefficients. This approach allows us to impose and investigate various windows in transverse momenta and rapidities, based on realistic kinematic configurations employed in current and future experimental studies at the LHC.
In particular, we will focus on the following two kinematic ranges.

%-----------------------------------------
\subsubsection{Standard LHC tagging}
\label{sssec:LHC_tagging}
%-----------------------------------------

We delve into the emission dynamics of both hadrons within the standard detection acceptances of CMS or ATLAS barrel calorimeters. 
Unlike the Mueller--Navelet channel, which extends into the end-cap region (as illustrated in panel ($a$) of Fig.~\ref{fig:detectors}), allowing for jet tagging with $|y_{\rm jet}| < 4.7$, hadron detection is confined to the barrel regions.
A realistic proxy for the rapidity range of hadron tags at the LHC can be derived from recent CMS analyses on $\Lambda_b$ particles~\cite{Chatrchyan:2012xg}, yielding $|y_{\Lambda_b}| < 2$. For our study, we adopt a slightly broader range, matching the coverage of the CMS barrel detector, $|y_{{\cal M}, {b}}| < 2.4$.

Following precedent studies~\cite{Celiberto:2017ptm,Celiberto:2020rxb,Celiberto:2021dzy}, we set a $10 < |\vec q_{T_{\cal M}}| / {\rm GeV} < 20$ window for the transverse momentum of the ${\cal M}$ hadron.
In contrast, we select a wider and disjoint transverse momentum range for the $b$-hadron, specifically $20 < |\vec q_{T_b}| / {\rm GeV} < 60$, as recently suggested in studies of high-energy $b$-flavored emissions~\cite{Celiberto:2021fdp,Celiberto:2024omj}. This selection ensures the robustness of our treatment within the VFNS framework, as energy scales consistently exceed the thresholds for DGLAP evolution of heavy-quark FFs (we refer to Section~\ref{ssec:PDFs_FFs} for further details).

On the one hand, the symmetric tagging in rapidity of both light and heavy particles enables us to rigorously apply our formalism in a well-defined regime, allowing for stringent tests of the $\NLLp$ hybrid factorization.
On the other hand, as highlighted in Ref.~\cite{Celiberto:2020wpk}, the use of disjoint transverse momentum intervals amplifies effects of additional, undetected gluon radiation by suppressing the Born contribution. This accentuates the signatures of the high-energy resummation over the fixed-order background.

Furthermore, asymmetric $|\vec q_T|$-windows mitigate potential instabilities arising in NLO calculations~\cite{Andersen:2001kta,Fontannaz:2001nq}, as well as violations of energy-momentum at NLL~\cite{Ducloue:2014koa}.
However, the achievable $\DY$ values with standard LHC tagging may not be sufficiently large to unambiguously discriminate between BFKL and fixed-order signals. This challenge can be addressed by exploring \emph{far-forward} emissions of one of the two particles, as proposed in Section~\ref{sssec:FPF-LHC_coincidence}.

%-----------------------------------------
\subsubsection{FPF + LHC coincidence}
\label{sssec:FPF-LHC_coincidence}
%-----------------------------------------

Expanding upon the standard detection regime outlined in Section~\ref{sssec:LHC_tagging}, we advocate for the simultaneous identification of an \emph{far-forward} particle (our choice falls into the  ${\cal M}$ hadron) alongside a more centrally located one (the $\Hb$ hadron, as depicted in panel ($b$) of Fig.~\ref{fig:detectors}).
Once the planned FPF~\cite{Anchordoqui:2021ghd,Feng:2022inv} becomes operational, this novel configuration might be achievable by coordinating FPF detectors with ATLAS. Integrating information from both ATLAS and the FPF hinges on the capability to trigger ATLAS events using far-forward signatures, necessitating precise timing protocols and influencing FPF detector design. Technical specifics regarding a possible FPF~$+$~ATLAS tight timing synchronization are detailed in Section~VI~E of Ref.~\cite{Anchordoqui:2021ghd}.

As a preliminary investigation, we examine the emission of a $\cal{M}$ meson within the far-forward range given by $5 < y_{\cal M} < 7$. While greater rapidities could be explored, we adopt a conservative approach, opting for a rapidity interval disjoint from and more forward than that accessible by end-caps of standard LHC detector. Investigations into larger rapidity ranges are deferred to future endeavors.
The meson is paired with a $b$-hadron detected by the LHC barrel calorimeter within the standard rapidity spectrum, $|y_{\cal H}| < 2.4$.
The transverse momenta of both final-state hadrons remain consistent with those specified in Section~\ref{sssec:LHC_tagging}.

The combined FPF~$+$~LHC tagging strategy yields a hybrid and markedly asymmetric range selection, presenting an excellent avenue for disentangling clear high-energy signals from collinear backgrounds.
However, as highlighted in Refs.~\cite{Celiberto:2020rxb,Bolognino:2021mrc} and mentioned previously, the joint detection of an far-forward particle alongside a central one results in an asymmetric configuration between the longitudinal momentum fractions of the corresponding incoming partons. This configuration significantly suppresses the contribution of undetected gluon radiation at LO and has a notable impact at NLO.

This kinematic constraint leads to an incomplete cancellation between virtual and real contributions from gluon emissions, resulting in the emergence of large threshold logarithms in the perturbative series.
Given that the BFKL framework encompasses the resummation of energy-type single logarithms while systematically neglecting threshold ones, we anticipate a partial degradation in the convergence of our resummed calculation in this FPF~$+$~LHC coincidence setup compared to the LHC standard scenario. Despite these challenges, these features motivate future developments aimed at incorporating the threshold resummation into our framework~\cite{Sterman:1986aj,Catani:1989ne,Catani:1996yz,Bonciani:2003nt,deFlorian:2005fzc,Ahrens:2009cxz,deFlorian:2012yg,Forte:2021wxe,Mukherjee:2006uu,Bolzoni:2006ky,Becher:2006nr,Becher:2007ty,Bonvini:2010tp,Bonvini:2018ixe,Ahmed:2014era,Banerjee:2018vvb,Duhr:2022cob,Shi:2021hwx,Wang:2022zdu}.

In conclusion, explorations via a FPF~$+$~LHC coincidence method offer an unparalleled opportunity for rigorous and in-depth examinations of strong interaction dynamics in the high-energy regime. In this regard, the advent of the planned FPF~\cite{Anchordoqui:2021ghd,Feng:2022inv} may complement the capabilities of the ATLAS detector, enabling us to (\emph{i}) assess the feasibility of precision analyses using hybrid high-energy and collinear factorization, and (\emph{ii}) explore potential commonalities among distinct resummation techniques.

%-----------------------------------------
\subsection{Rapidity-interval rates}
\label{ssec:Y_rates}
%-----------------------------------------

We delve into the analysis of $\DY$-rates for our studied processes, as illustrated in Figure~\ref{fig:process}. These distributions represent $\varphi$-summed, $\DY$-differential cross sections.
We note that clues of a \emph{natural stabilization} of these observables are strongly expected when a standard LHC tagging is considered, while they are awaited in a FPF~$+$~LHC coincidence setup. 
Observing such stability would not only be a further reliability test for the $\NLLp$ factorization, but it also would mark a milestone toward future precision analyses.

Stabilization effects can manifest at various levels. Firstly, the involve the ability to study $\DY$-distributions for all considered final states around the natural energy scales prescribed by kinematics. 
This prerequisite is essential for claiming evidence of stability, a feat unattainable when considering light hadrons and/or jets~\cite{Ducloue:2013bva,Caporale:2014gpa,Celiberto:2017ptm,Celiberto:2020wpk}. In such cases, large NLL contributions, of similar magnitude but opposite sign to their LL counterparts, can lead to unphysical $\DY$-distributions, sometimes even resulting in negative values for large $\DY$. Another sign of instability arises in the analysis of mean values of cosines of multiples of the azimuthal-angle distance, $\langle \cos (n \phi) \rangle$, which can exceed one.
Secondly, achieving a substantial reduction in the discrepancy between pure LL calculations and NLL-resummed ones would push natural stability to a higher level. 

To assess the impact of our resummed calculations on fixed-order predictions, we contrast $\LL$ and $\NLLp$ results with the corresponding ones derived using our $\HENLOp$ formula (Eq. \ref{Cn_HENLOp_MSb}), which effectively mimics the high-energy signal of a pure NLO computation. 
Given a dedicated numerical tool for higher-order perturbative calculations of cross sections for semi-hard hadroproduction of two identified hadrons is not yet available, our $\HENLOp$ approach remains the most valid and efficient method for a BFKL versus fixed-order comparison.

\begin{figure*}[t]
%\centering

   \includegraphics[scale=0.40,clip]{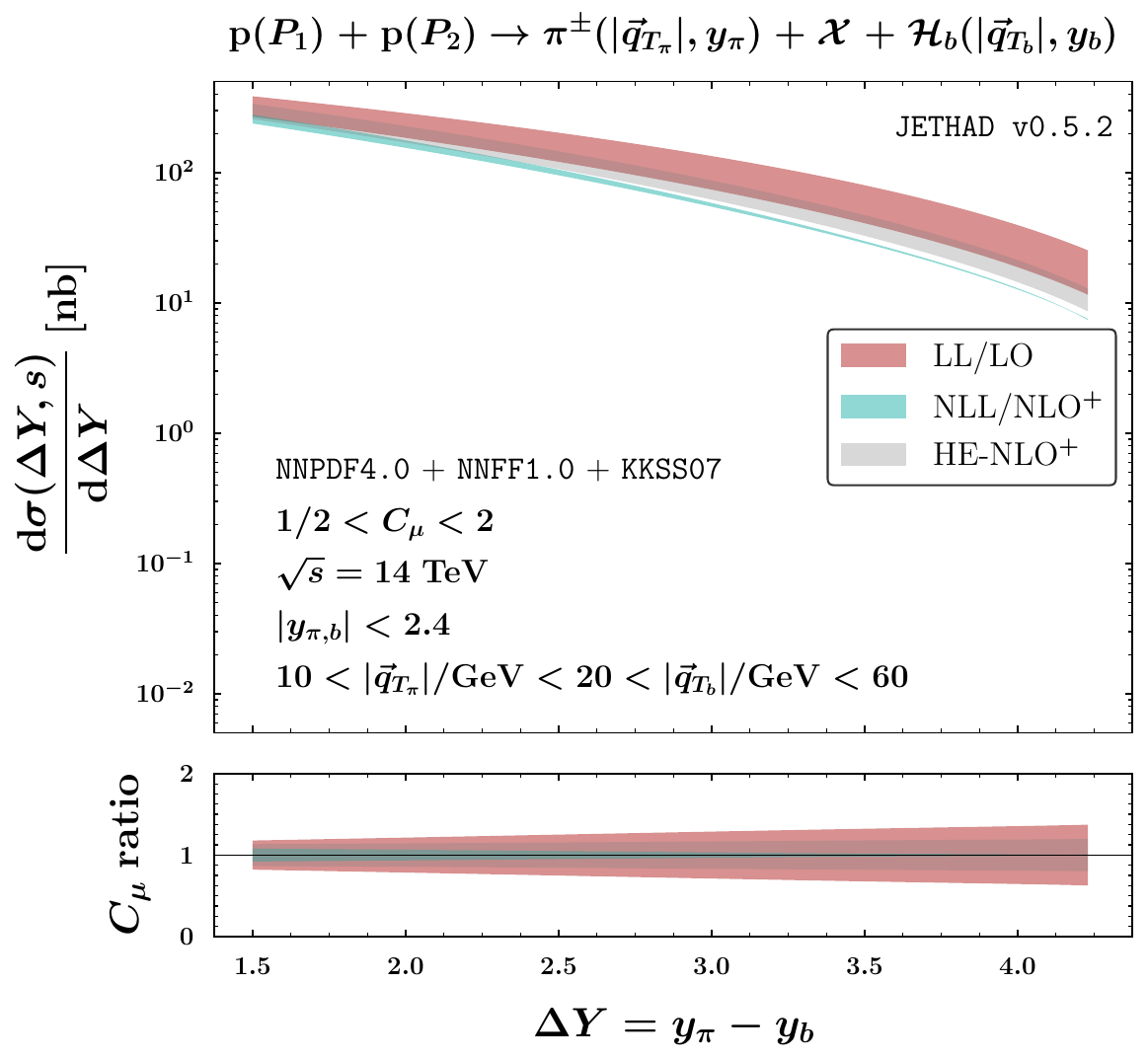}
   \hspace{0.35cm}
   \includegraphics[scale=0.40,clip]{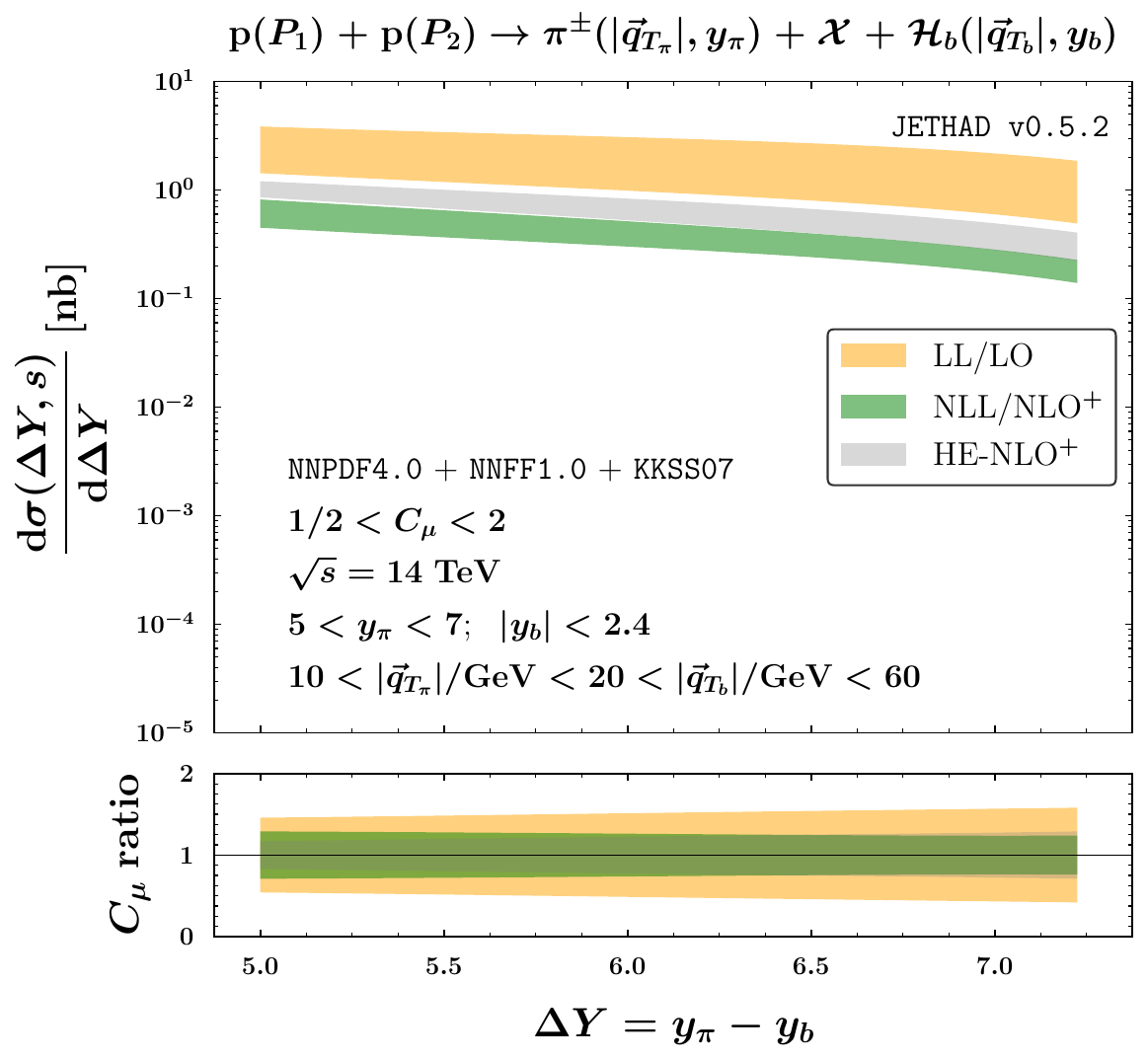}

   \vspace{0.50cm}

   \includegraphics[scale=0.40,clip]{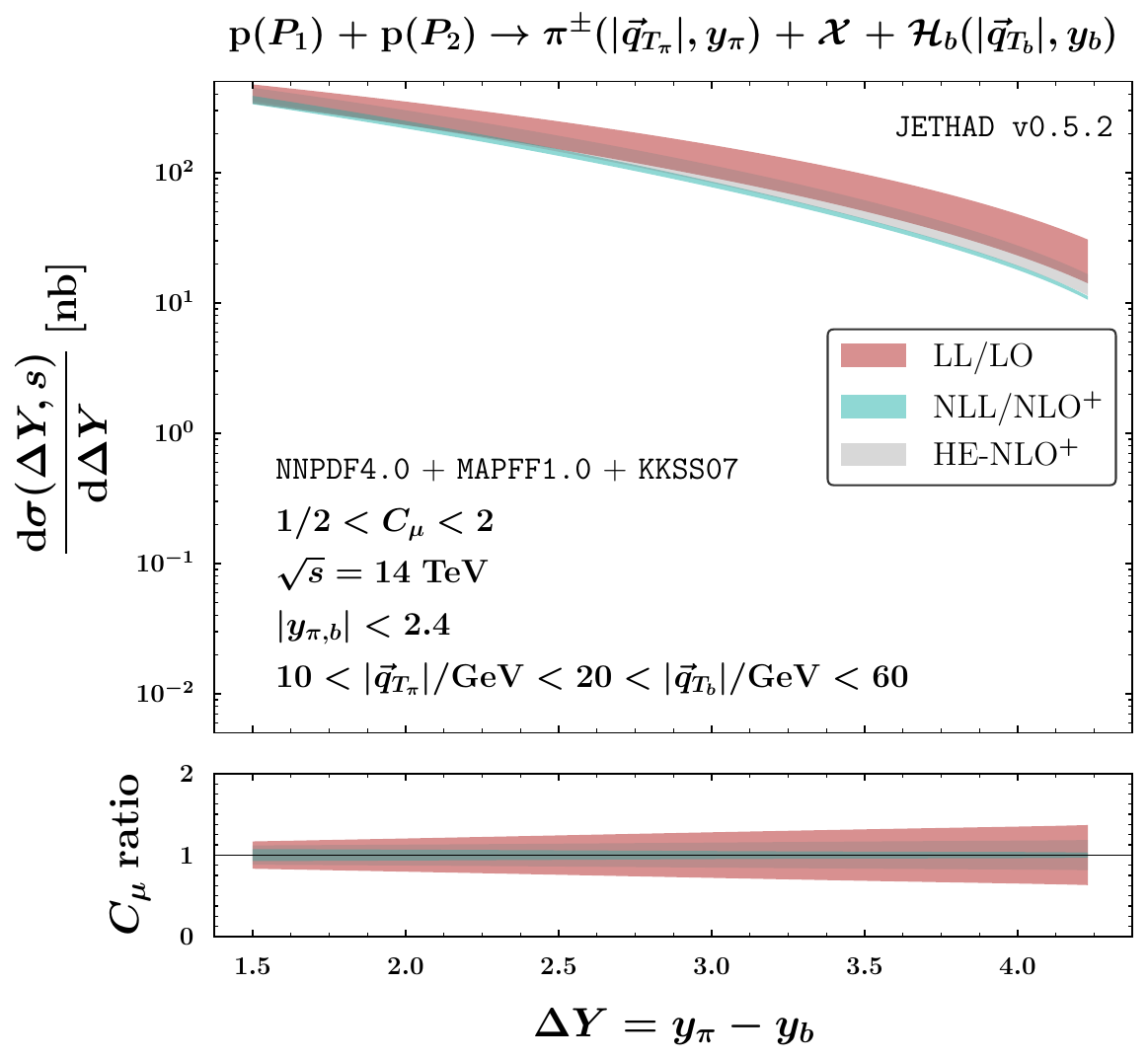}
   \hspace{0.35cm}
   \includegraphics[scale=0.40,clip]{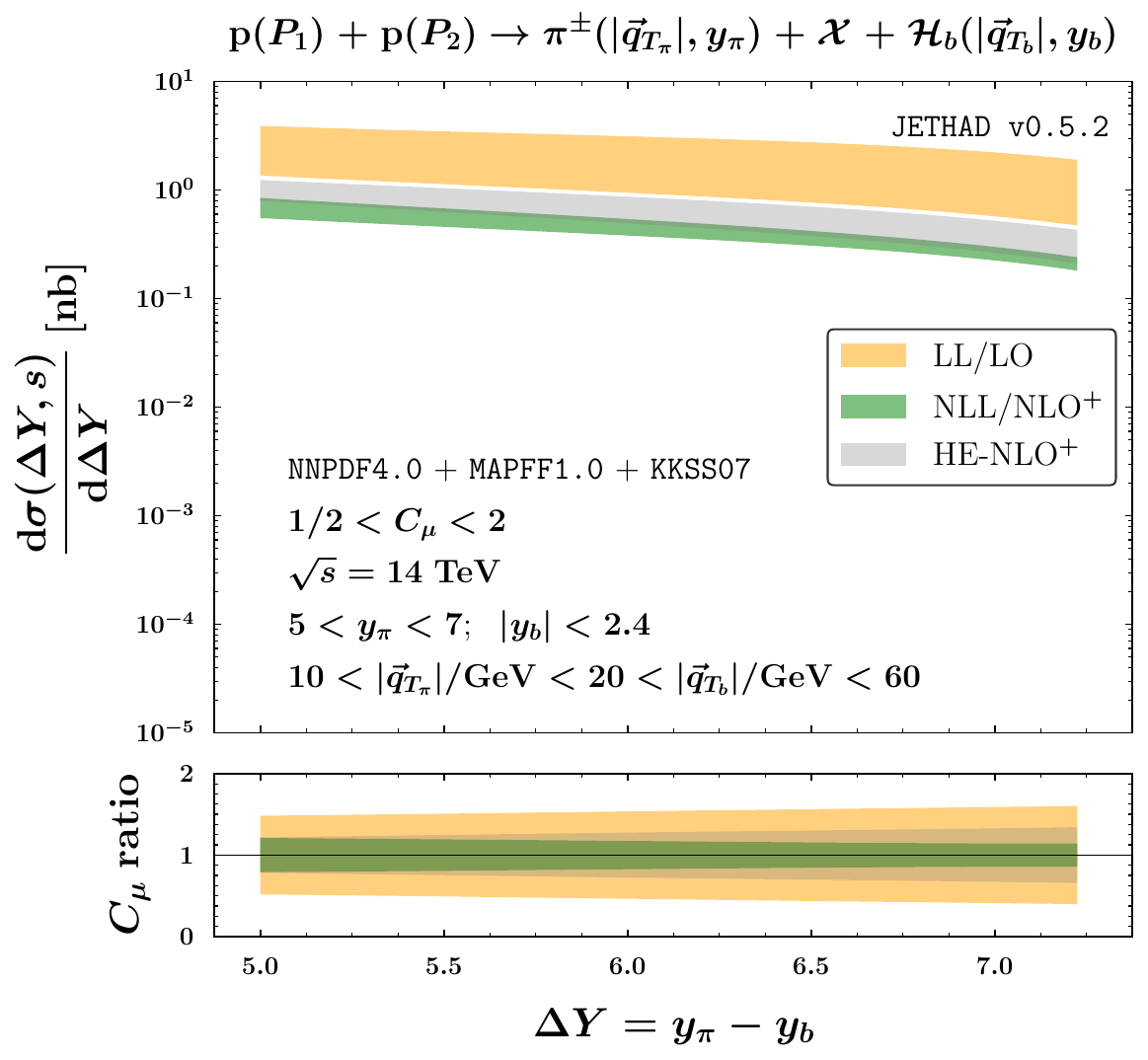}

\caption{Rapidity-interval rate for the inclusive $\pi^\pm$~$+$~$\Hb$ production at $\sqrt{s} = 14$ TeV LHC (left) and FPF~$+$~LHC (right).
Upper (lower) plots were obtained by using {\tt NNFF1.0} ({\tt MAPFF1.0}) pion NLO FFs together with {\tt KKSS07} $b$-hadron NLO FFs, and {\tt NNPDF4.0} NLO proton PDFs.}
\label{fig:DY_Pb}
\end{figure*}

\begin{figure*}[t]
%\centering

   \includegraphics[scale=0.40,clip]{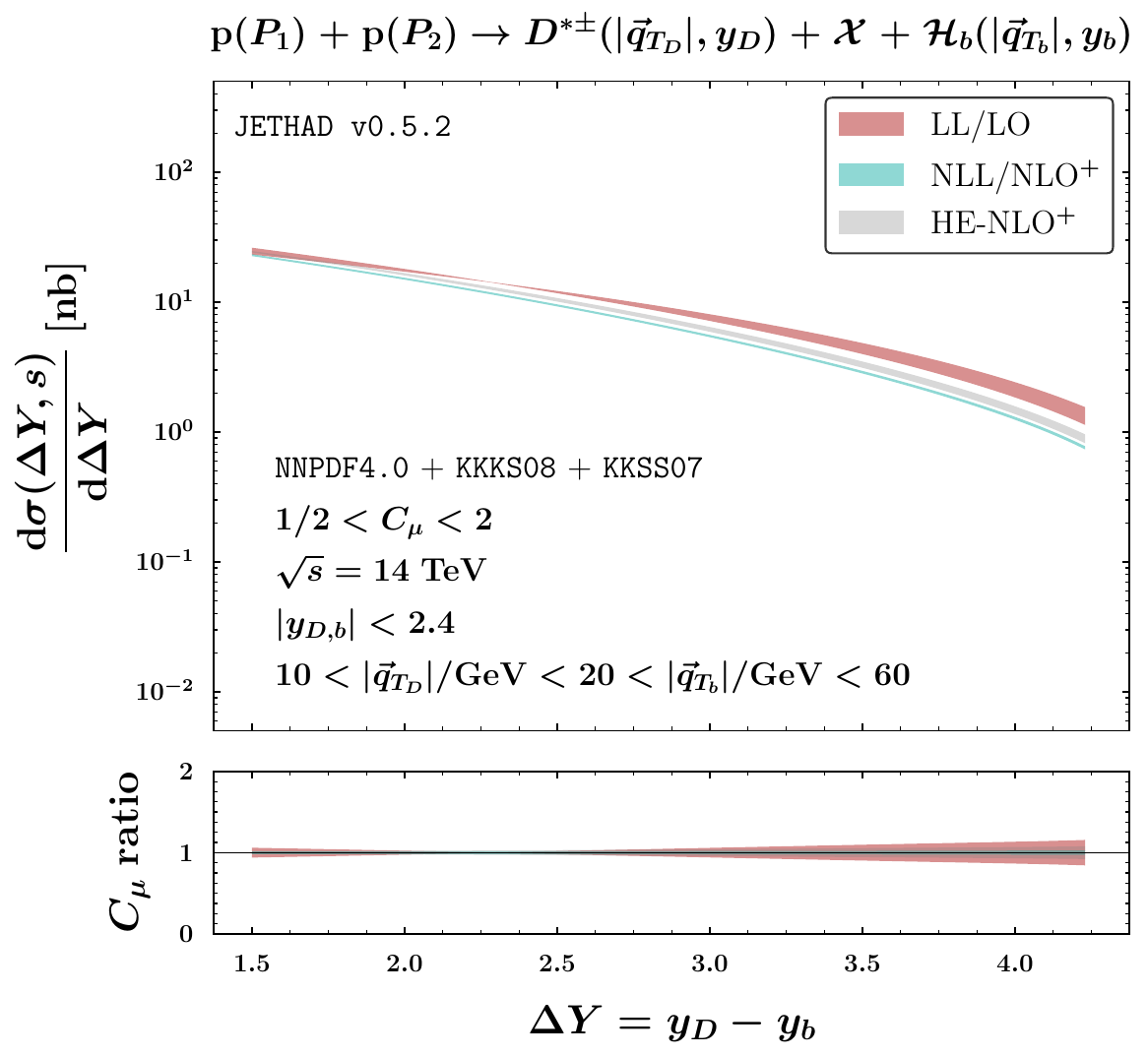}
   \hspace{0.35cm}
   \includegraphics[scale=0.40,clip]{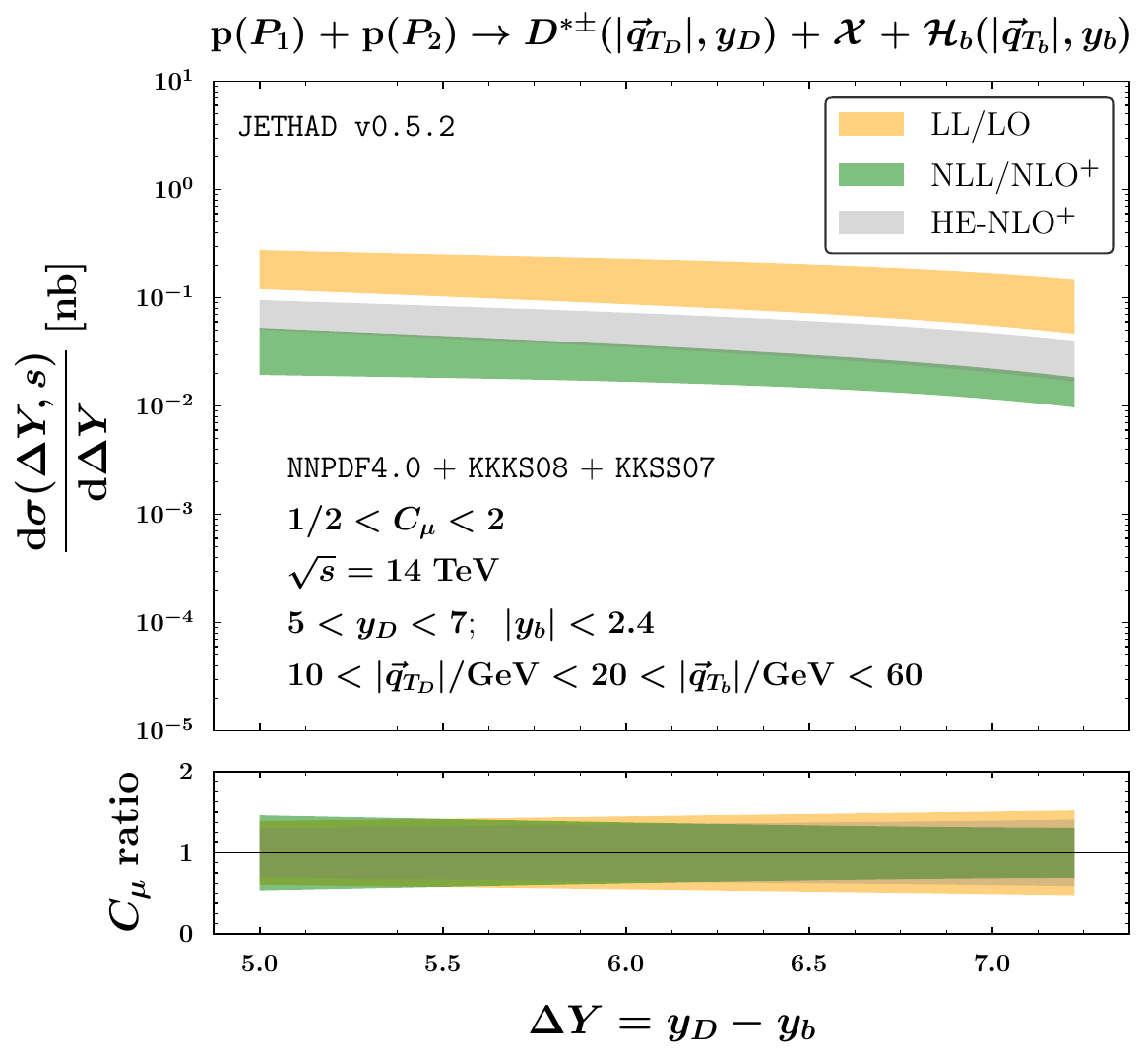}

\caption{Rapidity-interval rate for the inclusive $D^{*\pm}$~$+$~$\Hb$ production at $\sqrt{s} = 14$ TeV LHC (left) and FPF~$+$~LHC (right).
Plots were obtained by using {\tt KKKS08} $D$-meson NLO FFs together with {\tt KKSS07} $b$-hadron NLO FFs, and {\tt NNPDF4.0} NLO proton PDFs.}
\label{fig:DY_Db}
\end{figure*}

Plots of Fig.~\ref{fig:DY_Pb} depict the $\DY$-rate for inclusive $\pi^\pm \, + \, \Hb$ productions at the LHC (left panels) and at FPF~$+$~LHC (right panel).
Analogously, plots of Fig.~\ref{fig:DY_Db} refer to the inclusive $D{*^\pm} \, + \, \Hb$ detections.
Uncertainty bands are constructed considering the combined effect of MHOUs from energy-scale variation and numerical multi-dimensional integration over the final-state phase space, with the former being significantly dominant. 
Upon inspection, our $\DY$-distributions generally exhibit favorable statistics, lying in the range $10^{-1}$ to $5 \times 10^2$ nb.

The observed reflects the typical dynamics of our hybrid factorization. While the BFKL resummation predicts an increase in the partonic hard-scattering cross section with energy, its convolution with collinear PDFs and FFs leads to a decrease with $\DY$ for $\LL$, $\NLLp$, and $\NLLp$ results. This decrease is more pronounced for LHC kinematic configurations and appears smoother in the FPF~$+$~LHC case. The noncontiguous rapidity ranges covered by the FPF and current LHC detectors may contribute to this smoother shape, compensating for the increment with $\DY$ of the available phase space by the absence of detected events in the interval between $y_{\rm FPF}^{\rm min}$ and $y_{\rm LHC}^{\rm max}$.

A similar pattern was also noted in a complementary configuration, namely the CMS~$+$~CASTOR setup (see Fig.~10 of Ref.~\cite{Celiberto:2020wpk}). Results for rapidity-interval rates obtained with different pion FF parametrizations (upper versus lower panels of Fig.~\ref{fig:DY_Pb}) exhibit qualitatively similar trends, with their differences remaining significant, beyond a factor of three.

We observe the emergence of clear and natural stabilization effects in the high-energy series when standard LHC cuts are applied. 
The $\DY$-distributions for all the channels in Fig.~\ref{fig:process} feature $\NLLp$ bands that are partially or entirely nested within $\LL$ bands for small and moderate values of $\DY$. However, as the rapidity interval increases, $\NLL$ corrections become increasingly negative, causing $\NLLp$ predictions to become smaller than pure $\LL$ ones. $\NLL$ and $\HENLOp$ uncertainty bands are generally narrower than $\NLL$ and $\HENLOp$ ones. Furthermore, their width generally diminishes in the large $\DY$ range, where high-energy effects dominate over pure DGLAP ones. These observations are consistent with previous analyses on heavy-flavored emissions at CMS~\cite{Celiberto:2021dzy,Celiberto:2021fdp,Celiberto:2022dyf}, highlighting the convergence of the energy-resummed series thanks to the natural stabilizing effect of VFNS FFs, which accurately describe the hadronization mechanisms of the detected heavy-flavor species.

The high-energy stabilizing pattern is also evident in the FPF~$+$~LHC coincidence setup, although its effects are less pronounced (see right panels of Figs.~\ref{fig:phi_Pb_NNFF10} to~\ref{fig:phi_Db}). While our required condition for asserting evidence of stability is met, allowing for precision studies of cross sections around the natural energy scales provided by kinematics, it is notable that FPF~$+$~LHC $\NLLp$ predictions consistently remain below $\LL$ results. Additionally, $\NLLp$ uncertainty bands are narrower than $\LL$ ones, but slightly wider than $\NLLp$ bands for the same channels investigated in the standard LHC configurations. Furthermore, $\LL$ results consistently exceed $\HENLOp$ ones, while $\NLLp$ results are smaller. Although further dedicated studies are required to determine if the observed natural-stability signals degrade when the FPF rapidity acceptances are expanded beyond those imposed in our analysis, an explanation for the increased sensitivity of $\DY$-rates to the resummation accuracy can be provided based on our current understanding of the dynamics behind other resummation mechanisms.

We stress that the semi-hard nature of the final states considered leads to high energies but not necessarily to small-$x$ dynamics. This is particularly evident in the FPF~$+$~LHC coincidence setup, where the strongly asymmetric final-state rapidity ranges result in one of the two parton longitudinal fractions being consistently large, while the other takes more moderate values.
As highlighted in Section~\ref{sssec:FPF-LHC_coincidence}, our approach does not capture large-$x$ logarithms, whose inclusion would be performed via an appropriate resummation mechanism, the aforementioned threshold resummation. 
A significant finding from Ref.~\cite{Almeida:2009jt} on inclusive di-hadron detections in hadronic collisions is that incorporating NLL threshold resummation on top of pure NLO calculations leads to a notable increase in cross sections. Remarkably, this increase is comparable to the gap between our $\LL$ and $\NLLp$ high-energy predictions for $\DY$-distributions in FPF~$+$~LHC configurations.

Recent works~\cite{Liu:2020mpy,Shi:2021hwx} have demonstrated how NLL instabilities arising from forward hadron hadroproduction, described within the saturation framework~\cite{Stasto:2013cha}, can be substantially mitigated when threshold logarithms are incorporated into these calculations.
Considering these findings, we contend that the natural stability observed in our high-energy studies is not compromised by the adoption of FPF~$+$~LHC coincidence setups. It remains robust and provides a satisfactory description of $\DY$-rates at natural scales. The discrepancy between $\LL$ and $\NLLp$ predictions could come from those large-$x$, threshold logarithms, which are currently not accounted for by our hybrid factorization.
Incorporating the large-$x$ resummation represents a crucial step toward enhancing the description of the considered observables. Therefore, it should be pursued as the next logical step to evaluate the feasibility of precision studies of $\DY$-distributions at FPF~$+$~LHC.

The analysis presented in this section underscores that light-meson plus heavy-flavor production processes offer a promising avenue for stabilizing the high-energy resummation, as anticipated. $\DY$-distributions emerge as particularly promising observables for detecting signals of high-energy dynamics and potentially discriminating between BFKL-driven and fixed-order computations.
Further exploration of these distributions holds significant potential for delving into the interplay between high-energy QCD dynamics and other resummations, particularly the large-$x$ threshold effects. By probing these aspects in greater detail, we can gain deeper insights into the underlying mechanisms governing semi-hard phenomenology and refine our understanding of QCD dynamics in the high-energy regime.

%-----------------------------------------
\subsection{Angular multiplicities}
\label{ssec:phi_multiplicities}
%-----------------------------------------

Semi-hard phenomenology delves into observables that become increasingly sensitive to final-state rapidity intervals. When these observables are also differential in azimuthal angles, it exposes a core aspect of high-energy QCD. In the context of two-particle hadroproduction reactions, significant insights into the onset of BFKL dynamics emerge when large rapidity distances ($\DY$) enhance the weight of undetected gluons strongly ordered in rapidity. These gluon emissions, resummed as energy logarithms, induce a growing-with-$\DY$ decorrelation on the azimuthal plane of the outgoing particles.

\begin{figure*}[t]
%\centering

   \includegraphics[scale=0.40,clip]{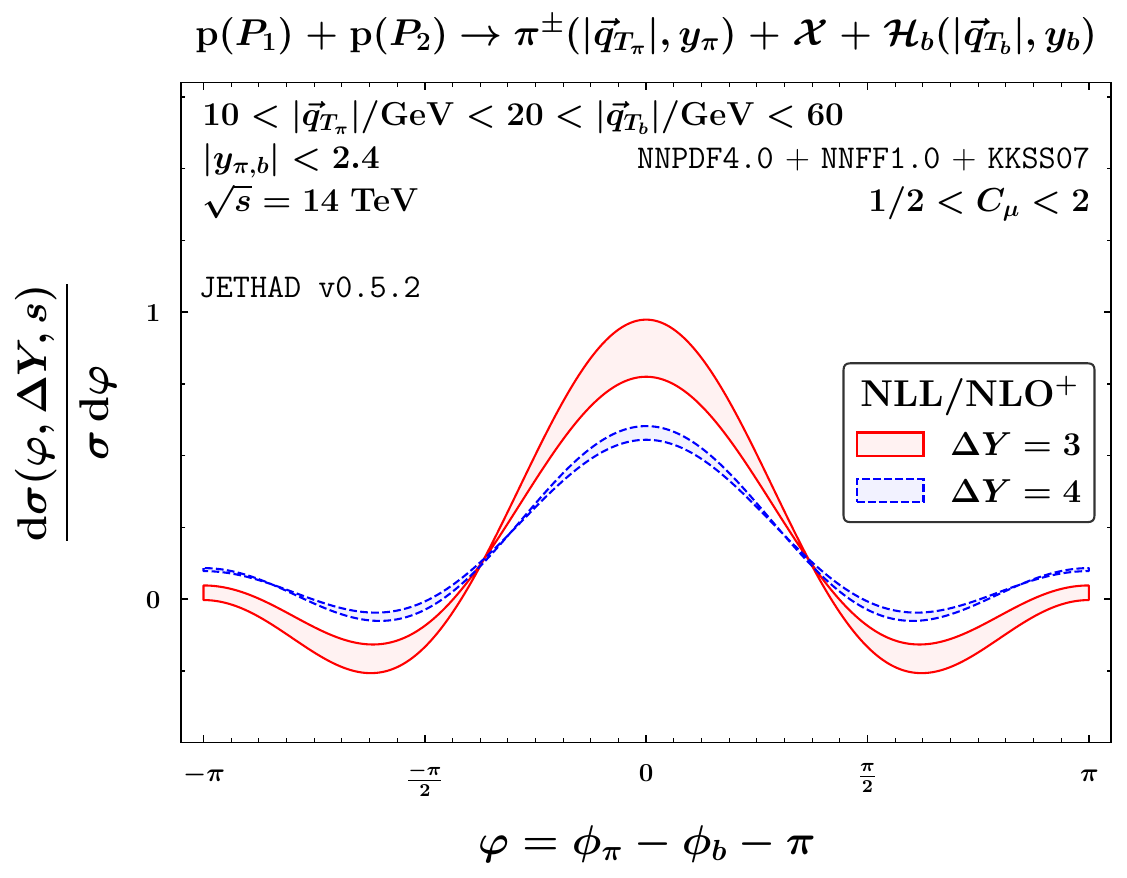}
   \hspace{0.35cm}
   \includegraphics[scale=0.40,clip]{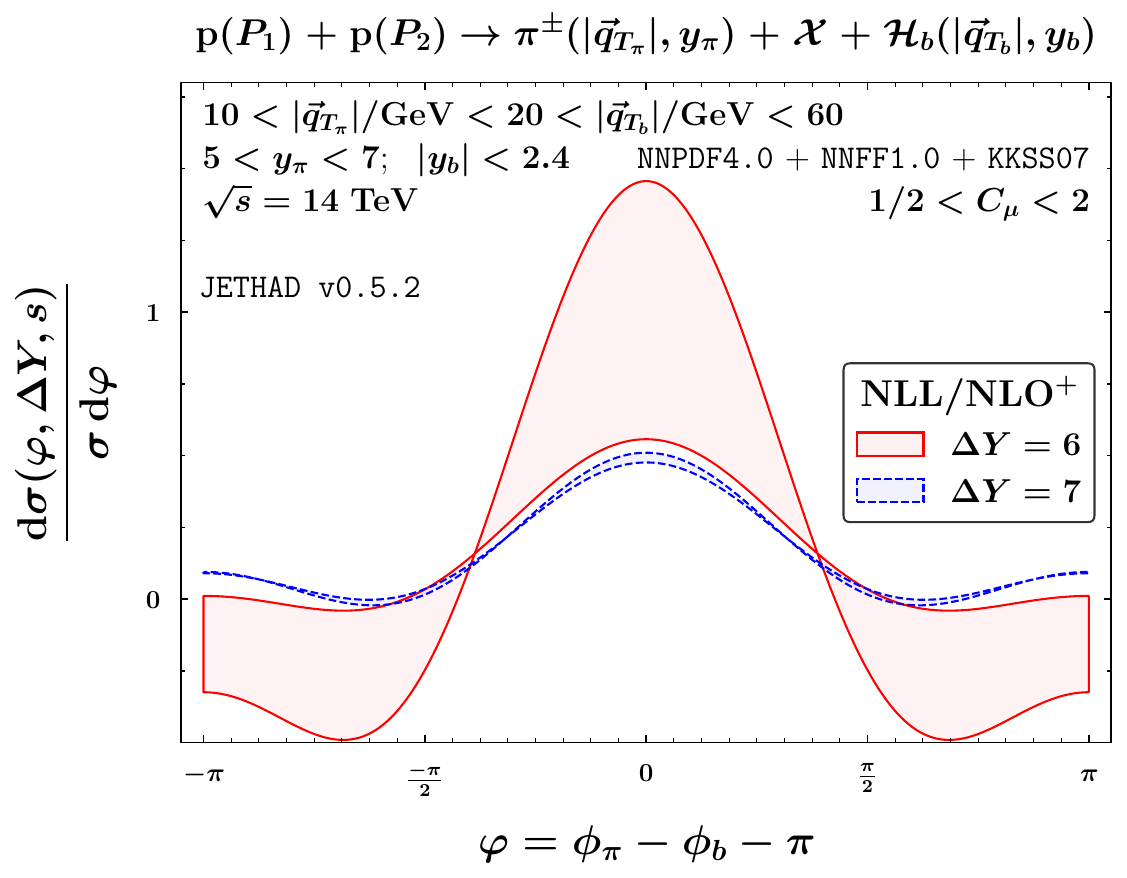}

   \vspace{0.50cm}

   \includegraphics[scale=0.40,clip]{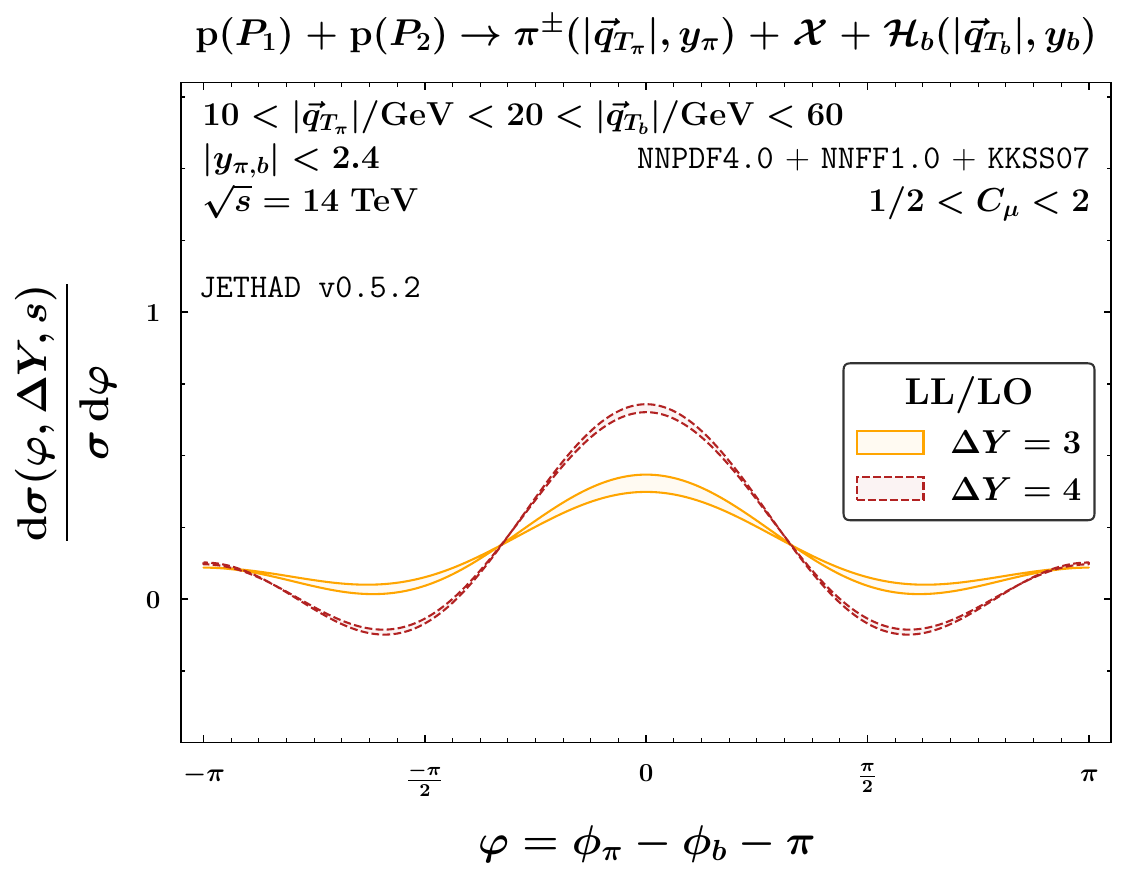}
   \hspace{0.35cm}
   \includegraphics[scale=0.40,clip]{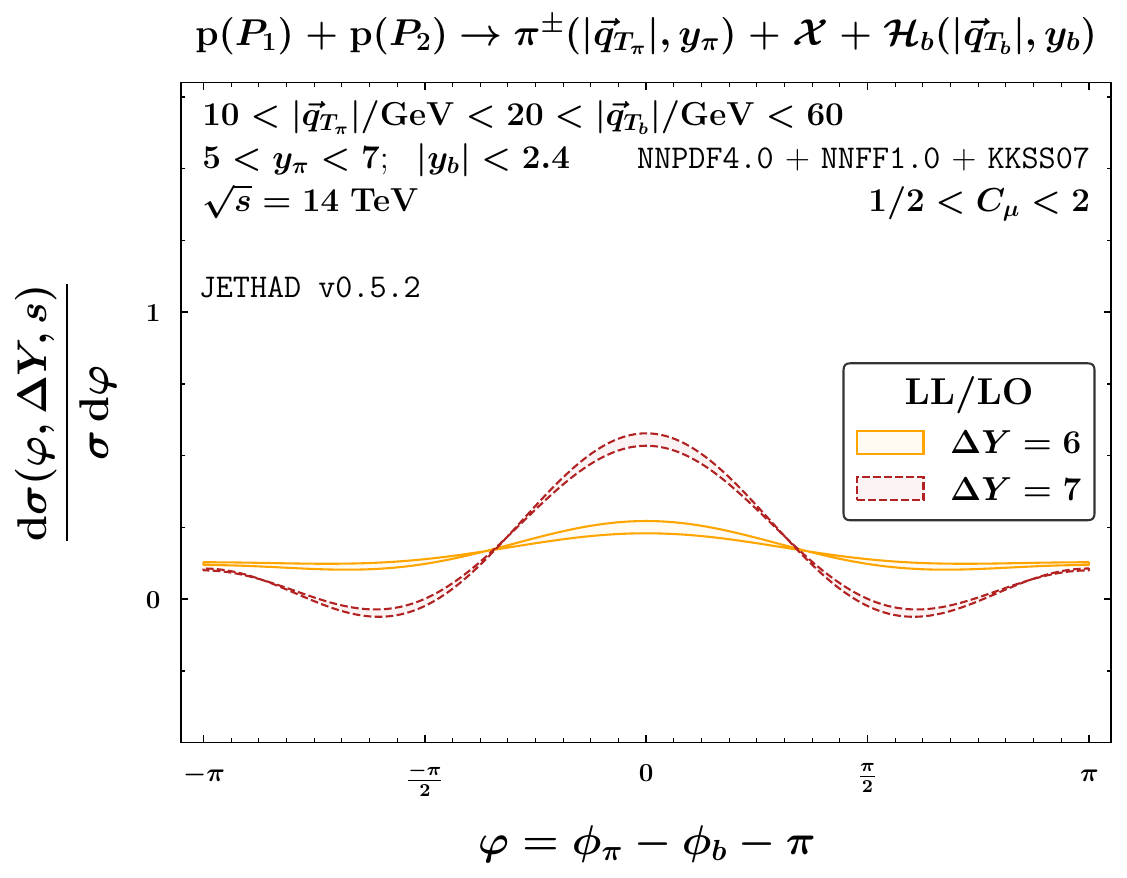}

\caption{Angular multiplicity for the inclusive $\pi^\pm$~$+$~$\Hb$ production at $\sqrt{s} = 14$ TeV LHC (left) and FPF~$+$~LHC (right).
Upper (lower) panels are for the $\NLLp$ ($\LL$) case.
Plots were obtained by using {\tt NNFF1.0} pion NLO FFs together with {\tt KKSS07} $b$-hadron NLO FFs, and {\tt NNPDF4.0} NLO proton PDFs.}
\label{fig:phi_Pb_NNFF10}
\end{figure*}

\begin{figure*}[t]
%\centering

   \includegraphics[scale=0.40,clip]{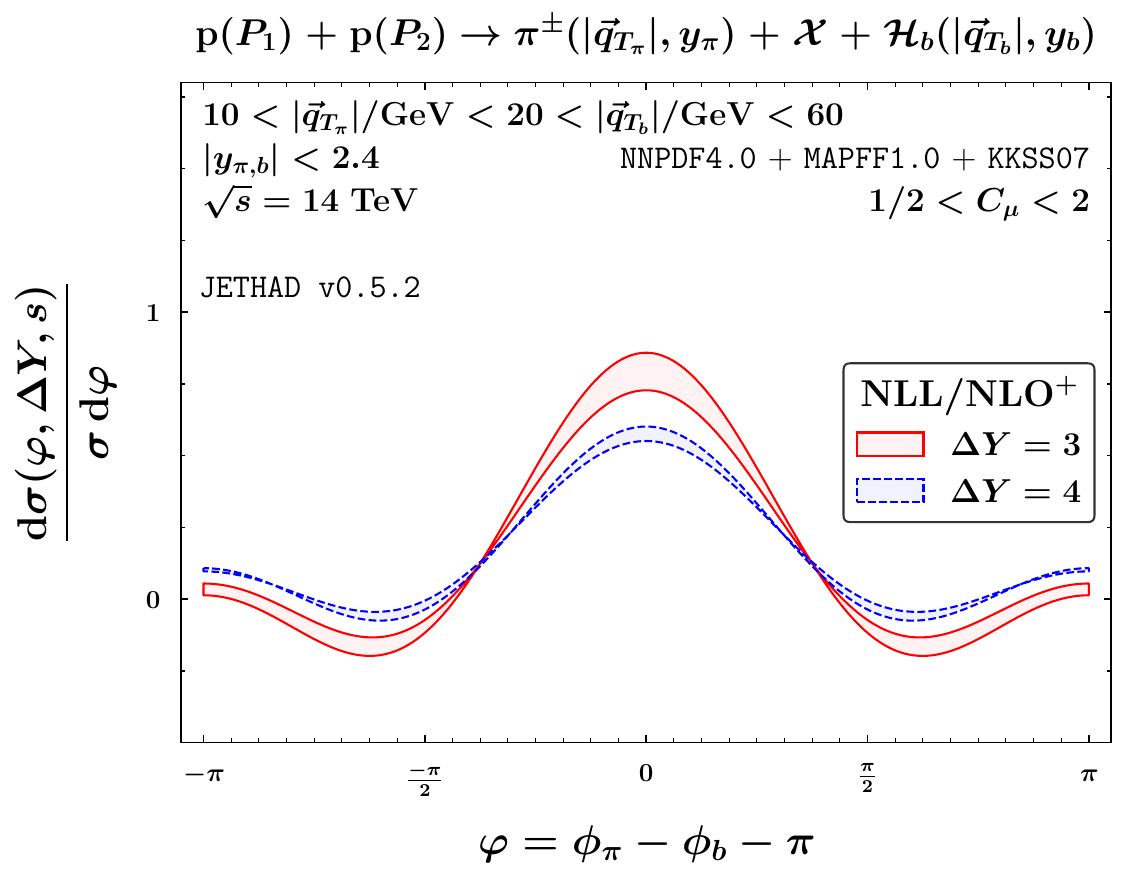}
   \hspace{0.35cm}
   \includegraphics[scale=0.40,clip]{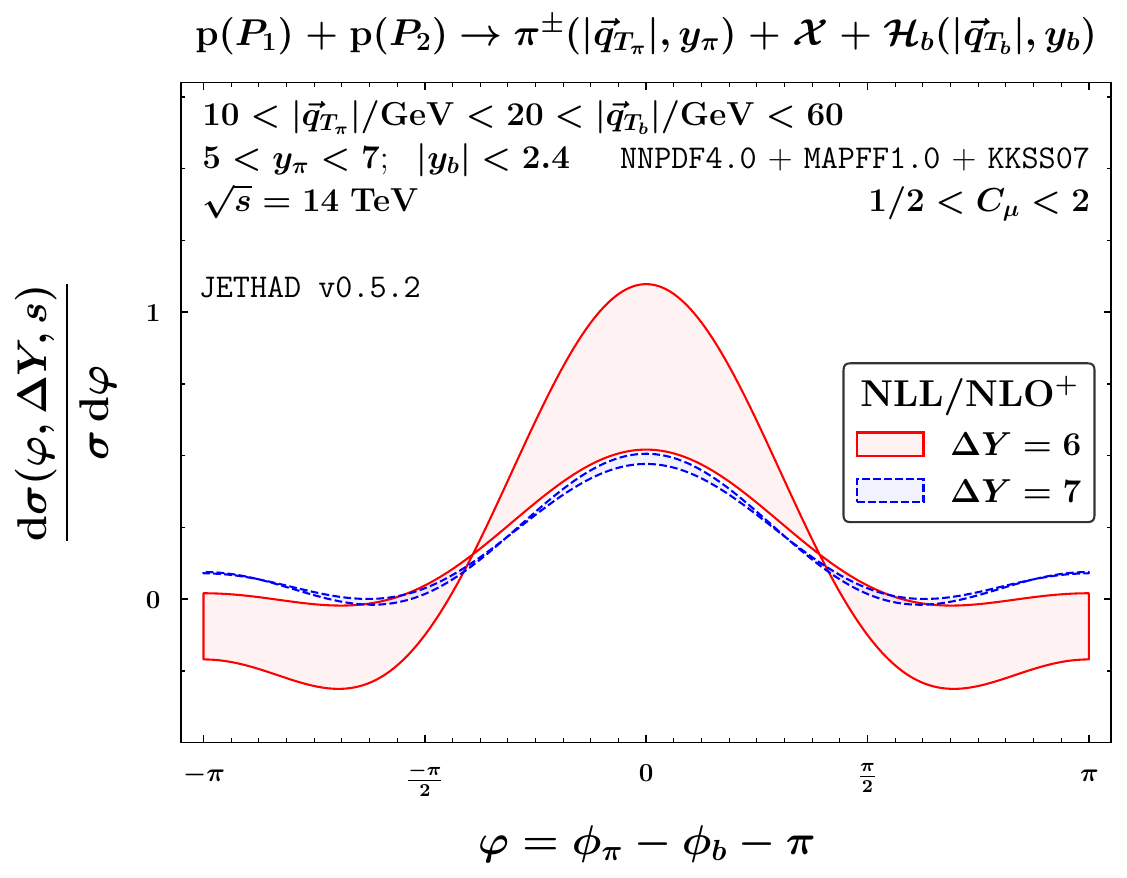}

   \vspace{0.50cm}

   \includegraphics[scale=0.40,clip]{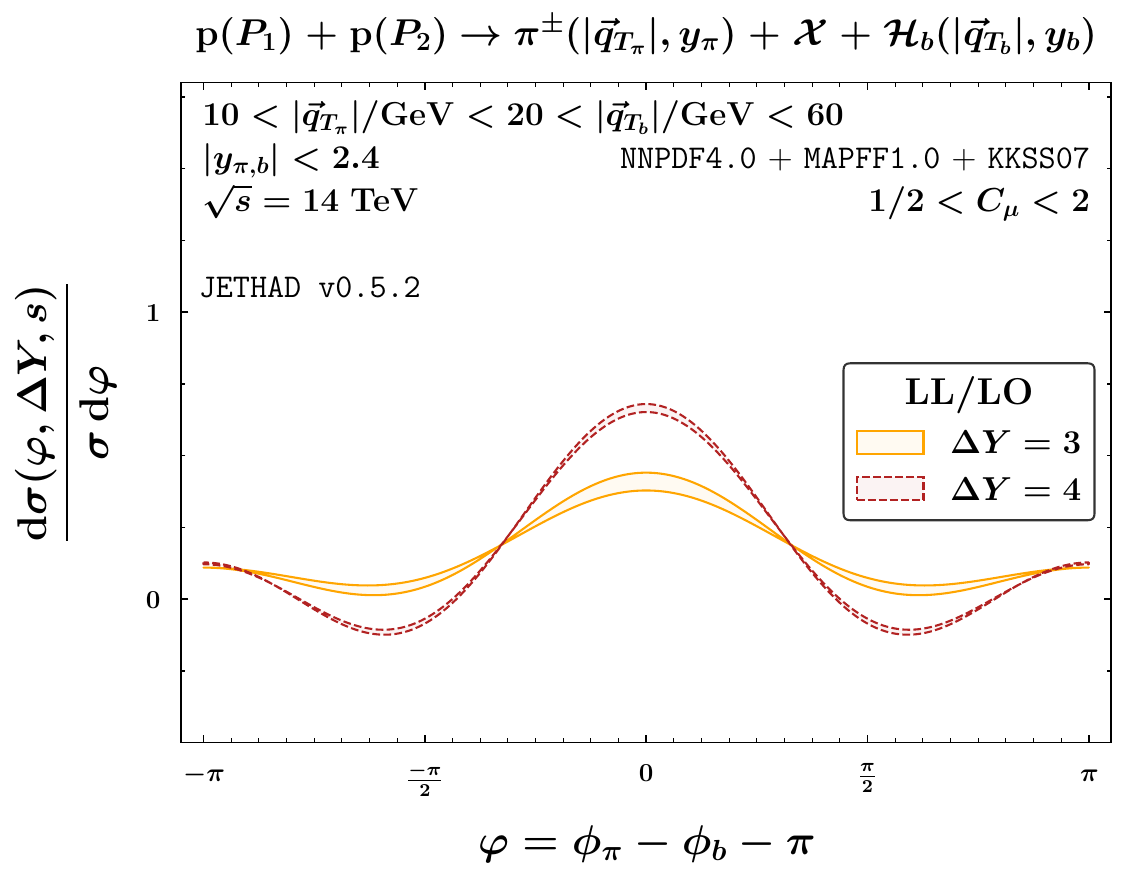}
   \hspace{0.35cm}
   \includegraphics[scale=0.40,clip]{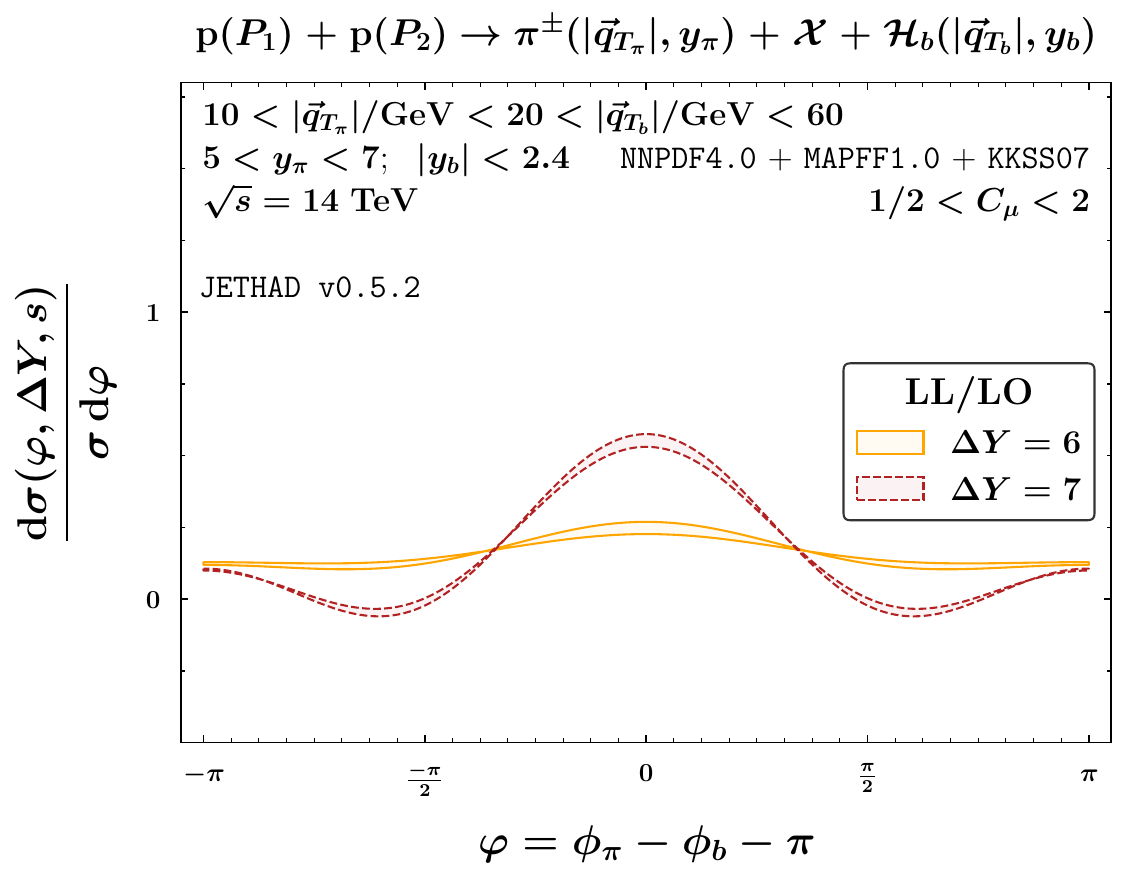}

\caption{Angular multiplicity for the inclusive $\pi^\pm$~$+$~$\Hb$ production at $\sqrt{s} = 14$ TeV LHC (left) and FPF~$+$~LHC (right).
Upper (lower) panels are for the $\NLLp$ ($\LL$) case.
Plots were obtained by using {\tt MAPFF1.0} pion NLO FFs together with {\tt KKSS07} $b$-hadron NLO FFs, and {\tt NNPDF4.0} NLO proton PDFs.}
\label{fig:phi_Pb_MAPFF10}
\end{figure*}

\begin{figure*}[t]
%\centering

   \includegraphics[scale=0.40,clip]{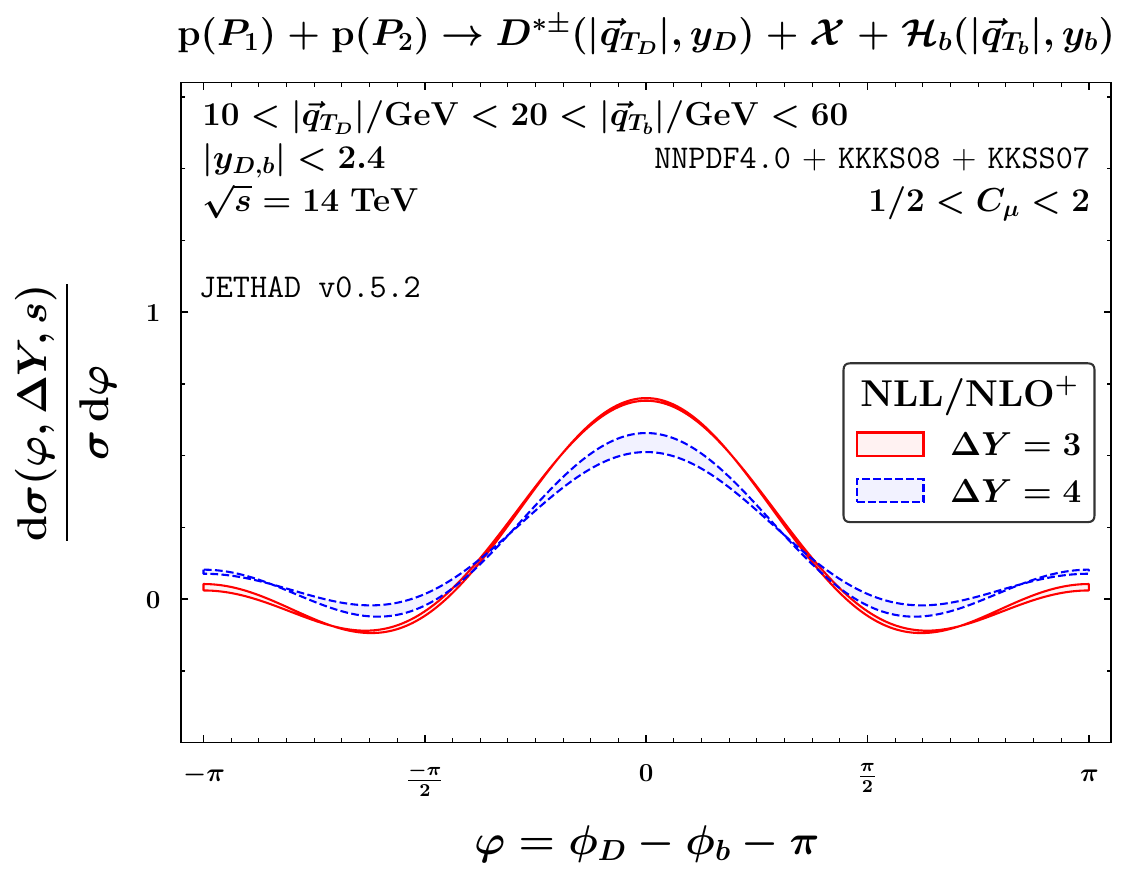}
   \hspace{0.35cm}
   \includegraphics[scale=0.40,clip]{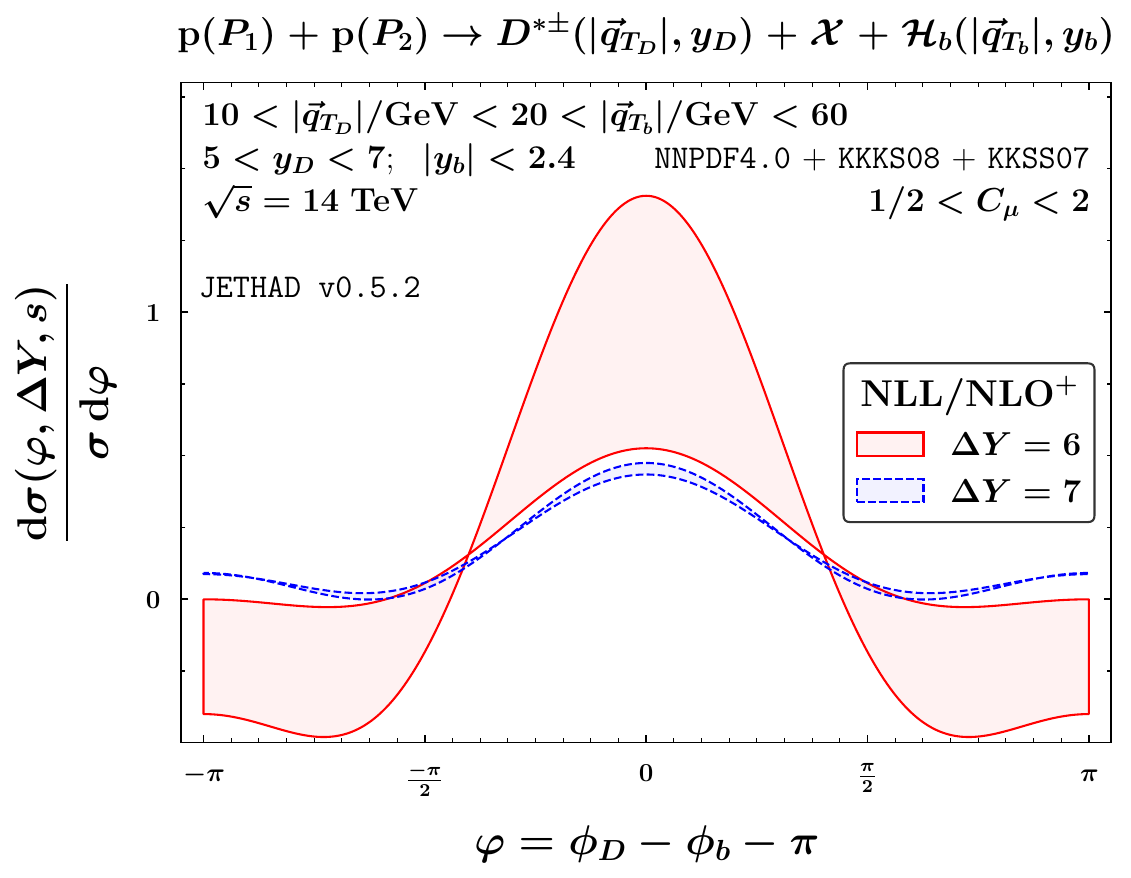}

\caption{Angular multiplicity for the inclusive $D^{*\pm}$~$+$~$\Hb$ production at $\sqrt{s} = 14$ TeV LHC (left) and FPF~$+$~LHC (right).
Upper (lower) panels are for the $\NLLp$ ($\LL$) case.
Plots were obtained by using {\tt KKKS08} $D$-meson NLO FFs together with {\tt KKSS07} $b$-hadron NLO FFs, and {\tt NNPDF4.0} NLO proton PDFs.}
\label{fig:phi_Db}
\end{figure*}

The angular decorrelation was initially observed through the $\DY$ dependence of cross-section azimuthal moment~\cite{Cerci:2008xv,Colferai:2010wu,Ducloue:2013bva,Ducloue:2013hia,Caporale:2014gpa}, defined as the ratios $R_{n0} \equiv C_n/C_0$ between a specific azimuthal coefficient ($C_{n \ge 1}$) and the $\varphi$-summed $C_0$ coefficient, which genuinely corresponds to the $\DY$ rate of Section~\ref{ssec:Y_rates}. 
The $R_{10}$ ratio effectively measures the azimuthal decorrelation between the two outgoing particles, akin to the mean value of $\langle \cos \varphi \rangle$. The $R_{n0}$ ratios represent the higher moments $\langle \cos (n \varphi) \rangle$, while further probes of BFKL were proposed through ratios between azimuthal moments, $R_{nm} \equiv C_n/C_m = \langle \cos (n \varphi) \rangle / \langle \cos (m \varphi) \rangle$, in earlier studies~\cite{Vera:2006un,Vera:2007kn}.

NLL-resummed predictions for angular correlations of Mueller--Navelet jets exhibited satisfactory agreement with LHC data at $\sqrt{s} = 7$ TeV, particularly for symmetric $|\vec q_T|$-windows~\cite{Ducloue:2013hia,Ducloue:2013bva,Caporale:2014gpa}. However, due to instabilities affecting the NLL series for processes involving the emission of two light-flavored particles, comparisons between theory and experiment had to be conducted at energy scales optimized via different procedures~\cite{Ducloue:2013hia,Caporale:2014gpa,Caporale:2015uva}.

Recent investigations into inclusive hadroproductions of heavy-flavored hadrons have revealed that the stabilizing effects associated with the use of heavy-flavor VFNS FFs are less pronounced when a heavy hadron is emitted alongside a light jet, as opposed to another heavy object~\cite{Celiberto:2021dzy,Celiberto:2021fdp}. Consequently, only a partial reduction in instabilities for azimuthal moments is observed.

As recently shown~\cite{Celiberto:2022dyf,Celiberto:2022gji}, starting from the angular coefficients, we can construct a more stable observable that contains signals of high-energy dynamics coming from all azimuthal modes. We refer to the \emph{angular multiplicity}
\begin{equation}
\label{angular_multiplicity}
 \frac{1}{\sigma}
 \frac{\drv \sigma}{\drv \varphi} (\varphi, \DY, s) = \frac{1}{2\pi} \left[ 1 + 2 \sum_{n = 1}^{\infty} \cos(n \varphi) \langle \, \cos(n \varphi) \rangle \right]
 = \frac{1}{2\pi} \left[ 1 + 2 \sum_{n = 1}^{\infty} \cos(n \varphi) \, R_{n0}(\DY, s) \right]
  \;,
\end{equation}
with $\varphi \equiv \phi_{\cal M} - \phi_{b} - \pi$.
This observable, initially proposed in Ref.~\cite{Marquet:2007xx} within the framework of Mueller--Navelet analyses, was further examined with NLL precision in Ref.~\cite{Ducloue:2013hia} for the same process.\footnote{The Fourier coefficients $R_{n0} (\DY,s)$ in Eq.~\eqref{angular_multiplicity} can be also employed to describe the azimuthal anisotropy of long-range correlations in heavy-ion collisions (see, \emph{e.g.}, Refs.~\cite{CMS:2017kcs,Mace:2019rtt,CMS:2019lin,Arslandok:2023utm,Ollitrault:2023wjk,ALICE:2023gyf,CMS:2023dse,CMS:2024krd} and references therein).} Exploring its characteristics holds significant advantages from both theoretical and experimental standpoints. On the one hand, it collects signals of high-energy dynamics from all azimuthal modes, rendering it one of the most robust observables for spotting BFKL effects.

As a normalized distribution, \emph{i.e.}, a \emph{multiplicity}, its sensitivity to uncertainties arising from uncertainties propagating from PDFs and/or FFs, as well as those from different replicas within the same set, is notably diminished. This allows for a focused examination of uncertainties intrinsic to high-energy resummation, facilitating stringent BFKL tests. On the other hand, experimental detector acceptances typically do not cover the entire ($2\pi$) azimuthal-angle range. Thus, comparing $\varphi$-dependent observables, like our azimuthal distribution, with experimental data is considerably more straightforward than for $R_{nm}$ correlation moments.

From a numerical standpoint, ensuring reliable predictions for our $\varphi$-distribution necessitates computing a large number of $C_n$ coefficients. We assessed the numerical stability of our calculations by progressively increasing the effective upper bound of the $n$ sum in Eq.~\eqref{angular_multiplicity}, achieving satisfactory convergence at $n_{\rm [bound]} \simeq 20$.

Results for angular multiplicities as functions of $\varphi$ and for two different sets of values for the rapidity interval, $\DY = 3, 4$ (LHC) or $\DY = 6, 7$ (FPF~$+$~LHC), are shown in Figs.~\ref{fig:phi_Pb_MAPFF10},~\ref{fig:phi_Pb_NNFF10}, and~\ref{fig:phi_Db}.
Panels in these figures are structured as follows. 
Upper (lowers) plots refer to $\NLLp$ ($\LL$) rates.
Standard LHC tagging and the FPF~$+$~LHC coincidence setups are used in left and right plots, respectively. Figures~\ref{fig:phi_Pb_NNFF10} and~\ref{fig:phi_Pb_MAPFF10} refer to pion channels described via {\tt NNFF1.0} and {\tt MAPFF1.0} pion collinear FFs, whereas Fig.~\ref{fig:phi_Db} show results for $D$-meson channels depicted by means of {\tt KKKS08} FF determinations.

The prominent feature shared among all the depicted multiplicities is the emergence of a peak centered at $\varphi = 0$, corresponding of the physical configuration where the $\cal{M}$ meson and the $\Hb$ hadron are emitted (almost) back-to-back. With increasing $\DY$, a characteristic trend emerges: the $\NLLp$ peak height decreases while its width expands. This trend stems from the heightened number of secondary gluons emitted with substantial rapidity separation, as predicted by the BFKL equation. Consequently, the correlation in the azimuthal plane between the two tagged particles diminishes, leading to a decrease in the number of nearly back-to-back events.

Conversely, $\LL$ distributions exhibit an opposing pattern: the peak grows while its width decreases with increasing $\DY$. This behavior generates from the connection between the strongly asymmetric transverse-momentum windows at which the two hadrons are tagged (see Section~\ref{ssec:final_state}) and the corresponding longitudinal-momentum fractions. 
This bring to a reduction in the combinations of these fractions for the given $\DY$. However, this results in an unphysical recorrelation pattern in the azimuthal plane for $\LL$ distributions, contradicting the expected loss of correlation due to the weight of rapidity-ordered gluons forming the inclusive system $\cal X$. The correct behavior is reinstated when full NLL corrections are considered.

Additionally, the size of uncertainty bands due to scale variations sensibly diminishes as $\DY$ increases.
This becomes strongly manifest in FPF~$+$~LHC selections. Multiplicities taken at lower reference values of the rapidity interval exhibit two symmetric minima at $|\varphi| \gtrsim \pi/2$, which extend to unphysical values beyond zero, whereas no negative values are observed for larger $\DY$ values. This indicates a natural stabilization of the high-energy series in the large rapidity-interval regime, as as expected.

Upon qualitative comparison of our predictions with corresponding results studied in other semi-hard channels, novel features emerge. 
As an example, angular distributions for our reactions appear less peaked compared to vector quarkonium $+$ jet distributions (Fig.~6 of Ref.~\cite{Celiberto:2022dyf}). 
They exhibit similarities with light-hadron or jet rates  (Figs.~17~and~18 of Ref.~\cite{Celiberto:2020wpk}).

In summary, multiplicities for $\cal{M}$-meson plus $b$-hadron productions within the hybrid factorization at $\NLLp$ allow for stringent tests of high-energy QCD dynamics. The natural stabilization of the high-energy resummation becomes significant in its expected applicability domain, particularly in the large $\DY$ sector.
This makes azimuthal distributions easily measurable in current LHC experimental configurations and future analyses facilitated by the FPF~$+$~LHC coincidence methods, this enhances our ability to conduct rigorous analyses of high-energy QCD dynamics.

%-----------------------------------------
\section{Toward precision studies of
high-energy QCD}
\label{sec:conclusions}
%-----------------------------------------

By employing the hybrid high-energy and collinear factorization approach at $\NLLp$, we have investigated the inclusive detection of pion or a $D$ meson, in association with a singly-bottomed hadron, at current LHC energies and kinematic configurations. Additionally, we explored configurations accessible via a FPF~$+$~LHC tight timing-coincidence setup.

Our analysis of distributions differential in the observed rapidity interval ($\DY$) or in the azimuthal-angle distance ($\phi$) between the two tagged objects has corroborated the remarkable property of \emph{natural stability} of the high-energy resummation. This feature, recently uncovered in the context of heavy-flavor studies in forward directions~\cite{Celiberto:2021dzy,Celiberto:2021fdp,Celiberto:2022dyf,Celiberto:2022keu,Celiberto:2023rzw}, enables a reliable description of the considered observables around the natural values of energy scales dictated by process kinematics. This stability is a prerequisite and an initial stride toward precision investigations of high-energy QCD through inclusive di-hadron system emissions in proton collisions.

We have demonstrated that $\DY$-rates effectively discriminate between the high-energy signal from the fixed-order background. On the other hand, $\varphi$-distributions exhibit robust stability in the large $\DY$ regime, offering a means to identify new and distinctive high-energy features.
The promising statistical outcomes of our observables in the FPF~$+$~LHC configuration underscore the interest of the FPF Community~\cite{Anchordoqui:2021ghd,Feng:2022inv} in exploring the intriguing prospect of enabling FPF and LHC detectors to operate in coincidence. Achieving this will necessitate extremely precise timing procedures, the technical feasibility of which should be actively pursued and complemented by positive feedback from the theoretical domain.

A striking conclusion from our recent investigation into the interplay between BFKL and DGLAP in inclusive semi-hard emissions of light jets and hadrons at the LHC~\cite{Celiberto:2020wpk} is the imperative for a \emph{multi-lateral} formalism. Such a framework would entail the simultaneous and consistent incorporation of several distinct resummation mechanisms, serving as a fundamental element for conducting precision studies of high-energy QCD. The sensitivity of FPF~$+$~LHC results presented in this study to both high-energy and threshold resummation underscores the urgency of developing such a unified description as a top priority in the medium-term future.

As a first prospect, we will complement our investigation on ${\cal M}$ mesons emitted in FPF-like kinematic configurations, where heavy particles are detected by LHC detectors, by examining the opposite configuration. In this setup, a far-forward heavy-flavored hadron is tagged at the FPF while another central object remains within LHC cuts. Subsequently, we will explore the high-energy behavior of observables sensitive to single inclusive emissions of heavy hadrons reconstructed by FPF detectors.
Our aim is to gain access to the proton content in far-forward (very low-$x$) regimes provided by FPF cuts. Here, our hybrid factorization framework could serve as a theoretical common basis for exploring production mechanisms and decays of heavy-flavored particles.

Then, by making use of estimates for differential cross sections of heavy-quark spectra at the particle-generated level, such as the ones provided via the {\tt FONLL} method~\cite{Cacciari:1998it,Cacciari:2001td,Cacciari:2012ny,Cacciari:2015fta}, it will be possible to make {\Jethad} work with parameters tuned for current high-energy hadron-hadron and lepton-hadron colliders which can produce heavy hadrons in forward directions of rapidity.

Mapping the proton structure in the very low-$x$ regime will hinge on a comprehensive exploration of the connections among different approaches. Specifically, we aim to investigate the interplay between our hybrid factorization, which permits the description of cross sections for single forward emissions in terms of a $\kappa_T$-factorization between off-shell matrix elements and the UGD and the ABF formalism which, as mentioned before, allowed us to obtain $\mbox{low-}x$ enhanced collinear PDFs (see also Section~6.1.2 of Ref.~\cite{Feng:2022inv}).

Then, as highlighted by model studies of leading-twist gluon TMD PDFs~\cite{Bacchetta:2020vty,Bacchetta:2024fci} (see also Refs.~\cite{Celiberto:2021zww,Bacchetta:2021oht,Bacchetta:2021lvw,Bacchetta:2021twk,Bacchetta:2022esb,Bacchetta:2022crh,Bacchetta:2022nyv,Celiberto:2022omz,Bacchetta:2023zir,Bacchetta:2024uxb,Hautmann:2017xtx,Hautmann:2017fcj,Monfared:2019uaj,Mukherjee:2023snp}), the distribution of linearly-polarized gluons can induce spin effects even in collisions of unpolarized hadrons. 
The latters, collectively known as the Boer--Mulders effect, were first observed in the case of quark polarization~\cite{Boer:1997nt,Bacchetta:2008xw,Barone:2008tn,Barone:2009hw}).
The gluon Boer--Mulders density can be readily accessed via inclusive emissions of heavy-flavored objects in hadron collisions, such as those that can be studied at FPFs (see Section 6.1.7 of Ref.~\cite{Feng:2022inv}).
Considering these insights, we aim to utilize FPF kinematic ranges as a tool to elucidate the connection between the BFKL UGD and the (un)polarized gluon TMDs.

Further progress will also hinge on connecting our program with NLO investigations of far-forward semi-inclusive emissions within the framework of gluon saturation~\cite{Gelis:2010nm,Kovchegov:2012mbw,Chirilli:2012jd,Boussarie:2014lxa,Benic:2016uku,Benic:2018hvb,Roy:2019hwr,Roy:2019cux,Beuf:2020dxl,Iancu:2021rup,Iancu:2023lel,vanHameren:2023oiq,Wallon:2023asa,Celiberto:2016yek,Celiberto:2016jse}. 
In this context, the influence of soft-gluon radiation on angular asymmetries in emissions of far-forward di-jet or di-hadron systems might be relevant~\cite{Hatta:2020bgy,Hatta:2021jcd,Caucal:2021ent,Caucal:2022ulg,Taels:2022tza,Fucilla:2022wcg}.
Higher-order saturation permits to explore of the (un)polarized gluon content of protons and nucleons at low-$x$~\cite{Kotko:2015ura,vanHameren:2016ftb,Altinoluk:2020qet,Altinoluk:2021ygv,Boussarie:2021ybe,Caucal:2023nci}. 
Studies in Refs.~\cite{Kang:2013hta,Ma:2014mri,Ma:2015sia,Ma:2018qvc,Stebel:2021bbn} consider into heavy-hadron emissions in proton-proton and proton-nucleus collisions while accounting for low-$x$ effects. 
Prospective inquiries of exclusive emission of heavy flavors in far-forward rapidity directions will unravel the connection between of our $\NLLp$ hybrid factorization NLO saturation~\cite{Mantysaari:2021ryb,Mantysaari:2022kdm}.

We view the analyses presented in this work as a significant step forward toward conducting precision studies of high-energy QCD. Our hybrid factorization framework, potentially enhanced via the integration of additional resummation techniques, offers a systematic approach to reducing uncertainties stemming from both perturbative calculations of high-energy scatterings and from collinear inputs. This serves a dual purpose: it provides a benchmark for SM measurements and establishes a shared foundation for the exploration of \ac{BSM} physics.

%-----------------------------------------
\section*{Acknowledgments}
%-----------------------------------------
\label{sec:acknowledgments}
\addcontentsline{toc}{section}{\nameref{sec:acknowledgments}}

The author is supported by the Atracci\'on de Talento Grant n. 2022-T1/TIC-24176 of the Comunidad Aut\'onoma de Madrid, Spain.
Feynman diagrams in Fig.~\ref{fig:process} were realized via the {\tt JaxoDraw 2.0} code~\cite{Binosi:2008ig}.

%\appendix
\begin{appendices}

%-----------------------------------------
\printacronyms
%-----------------------------------------

%-----------------------------------------
\setcounter{appcnt}{0}
\hypertarget{app:NLL_kernel}{
\section*{Appendix~A: High-energy kernel at NLL}}
\label{app:NLL_kernel}
%-----------------------------------------

The characteristic function encoded in the NLL correction to the high-energy kernel of Eq.~\eqref{chi_NLO} is
\begin{equation}
 \label{kernel_NLO}
 \bar \chi(n,\nu)\,=\, - \frac{1}{4}\left\{\frac{\pi^2 - 4}{3}\chi(n,\nu) - 6\zeta(3) - \frac{\drv^2 \chi}{\drv\nu^2} + \,2\,\phi(n,\nu) + \,2\,\phi(n,-\nu)
 \right.
\end{equation}
\[
 \left.
 +\frac{\pi^2\sinh(\pi\nu)}{2\,\nu\, \cosh^2(\pi\nu)}
 \left[
 \left(3+\left(1+\frac{n_f}{N_c^3}\right)\frac{11+12\nu^2}{16(1+\nu^2)}\right)
 \delta_{n0}
 -\left(1+\frac{n_f}{N_c^3}\right)\frac{1+4\nu^2}{32(1+\nu^2)}\delta_{n2}
\right]\right\} \, ,
\]
where
\begin{equation}
\label{kernel_NLO_phi}
 \phi(n,\nu)\,=\,-\int\limits_0^1 \drv x\,\frac{x^{-1/2+i\nu+n/2}}{1+x}\left\{\frac{1}{2}\left(\psi^\prime\left(\frac{n+1}{2}\right)-\zeta(2)\right)+\mbox{Li}_2(x)+\mbox{Li}_2(-x)\right.
\end{equation}
\[
\left.
 +\ln x\left[\psi(n+1)-\psi(1)+\ln(1+x)+\sum_{k=1}^\infty\frac{(-x)^k}{k+n}\right]+\sum_{k=1}^\infty\frac{x^k}{(k+n)^2}\left[1-(-1)^k\right]\right\}
\]
\[
 =\sum_{k=0}^\infty\frac{(-1)^{k+1}}{k+(n+1)/2+i\nu}\left\{\psi^\prime(k+n+1)-\psi^\prime(k+1)\right.
\]
\[
 \left.
 +(-1)^{k+1}\left[\Xi_\psi(k+n+1)+\Xi_\psi(k+1)\right]-\frac{\psi(k+n+1)-\psi(k+1)}{k+(n+1)/2+i\nu}\right\} \; ,
\]
\begin{equation}
\label{kernel_NLO_phi_beta_psi}
 \Xi_\psi(z)=\frac{1}{4}\left[\psi^\prime\left(\frac{z+1}{2}\right)
 -\psi^\prime\left(\frac{z}{2}\right)\right] \; ,
\end{equation}
and
\begin{equation}
\label{dilog}
\mbox{Li}_2(x) = - \int\limits_0^x \drv \zeta \,\frac{\ln(1-\zeta)}{\zeta} \; .
\end{equation}

%\clearpage

%-----------------------------------------
\setcounter{appcnt}{0}
\hypertarget{app:NLOHEF}{
\section*{Appendix~B: Forward-hadron emission function at NLO}}
\label{app:NLOHEF}
%-----------------------------------------

The NLO correction to the forward-hadron singly off-shell emission function can be written as~\cite{Ivanov:2012iv}

\begin{equation}
  \label{NLOHEF}
  \hat \J_h(n,\nu,|\vec q_T|,x)=
  \frac{1}{\pi}\sqrt{\frac{C_F}{C_A}}
  \left(|\vec q_T|^2\right)^{i\nu-\frac{1}{2}}
  \int_{x}^1\frac{\drv \zeta}{\zeta}
  \int_{\frac{x}{\zeta}}^1\frac{\drv \vartheta}{\vartheta}
  \left(\frac{\zeta\vartheta}{x}\right)^{2i\nu-1}
\end{equation}
  \[ \times \,
  \left[
  \frac{C_A}{C_F}f_g(\zeta)D_g^h\left(\frac{x}{\zeta\vartheta}\right){\cal C}_{gg}
  \left(\zeta,\vartheta\right)+\sum_{i=q\bar q}f_i(\zeta)D_i^h
  \left(\frac{x}{\zeta\vartheta}
  \right){\cal C}_{qq}\left(\zeta,\vartheta\right)
  \right.
  \]
  \[ + \,
  \left.D_g^h\left(\frac{x}{\zeta\vartheta}\right)
  \sum_{i=q\bar q}f_i(\zeta){\cal C}_{qg}
  \left(\zeta,\vartheta\right)+\frac{C_A}{C_F}f_g(\zeta)\sum_{i=q\bar q}D_i^h
  \left(\frac{x}{x\vartheta}\right){\cal C}_{gq}\left(\zeta,\vartheta\right)
  \right]\, .
  \]
Here, the ${\cal C}_{ij}$ parton coefficients read
\begin{equation}
\stepcounter{appcnt}
\label{Cgg_hadron}
 {\cal C}_{gg}\left(\zeta,\vartheta\right) =  2 P_{gg}(\vartheta)\left(1+\vartheta^{-2\gamma}\right)
 \ln \left( \frac {|\vec q_T| \zeta \vartheta}{\mu_F x}\right)
 - \beta_0 \ln \left( \frac {|\vec q_T| \zeta \vartheta}
 {\mu_R x}\right)
\end{equation}
\[
 + \, \delta(1-\vartheta)\left[C_A \ln\left(\frac{s_0 \, \zeta^2}{|\vec q_T|^2 \,
 x^2 }\right) \chi(n,\gamma)
 - C_A\left(\frac{67}{18}-\frac{\pi^2}{2}\right)+\frac{5}{9}n_f
 \right.
\]
\[
 \left.
 +\frac{C_A}{2}\left(\psi^\prime\left(1+\gamma+\frac{n}{2}\right)
 -\psi^\prime\left(\frac{n}{2}-\gamma\right)
 -\chi^2(n,\gamma)\right) \right]
 + \, C_A \left(\frac{1}{\vartheta}+\frac{1}{(1-\vartheta)_+}-2+\vartheta\bar\vartheta\right)
\]
\[
 \times \, \left(\chi(n,\gamma)(1+\vartheta^{-2\gamma})-2(1+2\vartheta^{-2\gamma})\ln\vartheta
 +\frac{\bar \vartheta^2}{\vartheta^2}{\cal I}_2\right)
\]
\[
 + \, 2 \, C_A (1+\vartheta^{-2\gamma})
 \left(\left(\frac{1}{\vartheta}-2+\vartheta\bar\vartheta\right) \ln\bar\vartheta
 +\left(\frac{\ln(1-\vartheta)}{1-\vartheta}\right)_+\right) \ ,
\]

\begin{equation}
\stepcounter{appcnt}
\label{Cgq_hadron}
 {\cal C}_{gq}\left(\zeta,\vartheta\right) = 
 2 P_{qg}(\vartheta)\left(\frac{C_F}{C_A}+\vartheta^{-2\gamma}\right)\ln \left( \frac {|\vec q_T| \zeta \vartheta}{\mu_F x}\right)
\end{equation}
\[
 + \, 2 \, \vartheta \bar\vartheta \, T_R \, \left(\frac{C_F}{C_A}+\vartheta^{-2\gamma}\right)+\, P_{qg}(\vartheta)\, \left(\frac{C_F}{C_A}\, \chi(n,\gamma)+2 \vartheta^{-2\gamma}\,\ln\frac{\bar\vartheta}{\vartheta} + \frac{\bar \vartheta}{\vartheta}{\cal I}_3\right) \ ,
\]

\begin{equation}
\stepcounter{appcnt}
\label{qg}
 {\cal C}_{qg}\left(\zeta,\vartheta\right) =  2 P_{gq}(\vartheta)\left(\frac{C_A}{C_F}+\vartheta^{-2\gamma}\right)\ln \left( \frac {|\vec q_T| \zeta \vartheta}{\mu_F x}\right)
\end{equation}
\[
 + \vartheta\left(C_F\vartheta^{-2\gamma}+C_A\right) + \, \frac{1+\bar \vartheta^2}{\vartheta}\left[C_F\vartheta^{-2\gamma}(\chi(n,\gamma)-2\ln\vartheta)+2C_A\ln\frac{\bar \vartheta}{\vartheta} + \frac{\bar \vartheta}{\vartheta}{\cal I}_1\right] \ ,
\]
and
\begin{equation}
\stepcounter{appcnt}
\label{Cqq_hadron}
 {\cal C}_{qq}\left(x,\vartheta\right) = 
 2 P_{qq}(\vartheta)\left(1+\vartheta^{-2\gamma}\right)\ln \left( \frac {|\vec q_T| \zeta \vartheta}{\mu_F x}\right) 
 - \beta_0 \ln \left(\frac {|\vec q_T| \zeta \vartheta}{\mu_R x}\right)
\end{equation}
\[
 + \, \delta(1-\vartheta)\left[- C_A \ln\left(\frac{s_0 \, \zeta^2}{|\vec q_T|^2 \, x^2}\right) \chi(n,\gamma)+ C_A\left(\frac{85}{18}+\frac{\pi^2}{2}\right)-\frac{5}{9}n_f - 8\, C_F \right.
\]
\[
 \left. +\frac{C_A}{2}\left(\psi^\prime\left(1+\gamma+\frac{n}{2}\right)-\psi^\prime\left(\frac{n}{2}-\gamma\right)-\chi^2(n,\gamma)\right) \right] + \, C_F \,\bar \vartheta\,(1+\vartheta^{-2\gamma})
\]
\[
 +\left(1+\vartheta^2\right)\left[C_A (1+\vartheta^{-2\gamma})\frac{\chi(n,\gamma)}{2(1-\vartheta )_+}+\left(C_A-2\, C_F(1+\vartheta^{-2\gamma})\right)\frac{\ln \vartheta}{1-\vartheta}\right]
\]
\[
 +\, \left(C_F-\frac{C_A}{2}\right)\left(1+\vartheta^2\right)\left[2(1+\vartheta^{-2\gamma})\left(\frac{\ln (1-\vartheta)}{1-\vartheta}\right)_+ + \frac{\bar \vartheta}{\vartheta^2}{\cal I}_2\right] \; ,
\]
with $T_R = 1/2$.
The $s_0$ scale is an addition energy scale that we set to $s_0 = \mu_C$.
Furthermore, one has $\bar \vartheta \equiv 1 - \vartheta$ and $\gamma \equiv - \frac{1}{2} + i \nu$. The LO DGLAP kernels $P_{i j}(\vartheta)$ are given by
\begin{eqnarray}
\stepcounter{appcnt}
\label{DGLAP_kernels}
 %\nonumber
 P_{gq}(z)&=&C_F\frac{1+(1-z)^2}{z} \; , \\ \nonumber
 P_{qg}(z)&=&T_R\left[z^2+(1-z)^2\right]\; , \\ \nonumber
 P_{qq}(z)&=&C_F\left( \frac{1+z^2}{1-z} \right)_+= C_F\left[ \frac{1+z^2}{(1-z)_+} +{3\over 2}\delta(1-z)\right]\; , \\ \nonumber
 P_{gg}(z)&=&2C_A\left[\frac{1}{(1-z)_+} +\frac{1}{z} -2+z(1-z)\right]+\left({11\over 6}C_A-\frac{n_f}{3}\right)\delta(1-z) \; ,
\end{eqnarray}
whereas the ${\cal I}_{2,1,3}$ functions read
\begin{equation}
\stepcounter{appcnt}
\label{I2}
{\cal I}_2=
\frac{\vartheta^2}{\bar \vartheta^2}\left[
\vartheta\left(\frac{{}_2F_1(1,1+\gamma-\frac{n}{2},2+\gamma-\frac{n}{2},\vartheta)}
{\frac{n}{2}-\gamma-1}-
\frac{{}_2F_1(1,1+\gamma+\frac{n}{2},2+\gamma+\frac{n}{2},\vartheta)}{\frac{n}{2}+
\gamma+1}\right)\right.
\end{equation}
\[
 \stepcounter{appcnt}
 \left.
 +\vartheta^{-2\gamma}\left(\frac{{}_2F_1(1,-\gamma-\frac{n}{2},1-\gamma-\frac{n}{2},\vartheta)}{\frac{n}{2}+\gamma}-\frac{{}_2F_1(1,-\gamma+\frac{n}{2},1-\gamma+\frac{n}{2},\vartheta)}{\frac{n}{2} -\gamma}\right)
\right.
\]
\[
 \left.
+\left(1+\vartheta^{-2\gamma}\right)\left(\chi(n,\gamma)-2\ln \bar \vartheta \right)+2\ln{\vartheta}\right] \; ,
\]
\begin{equation}
\stepcounter{appcnt}
\label{I1}
 {\cal I}_1=\frac{\bar \vartheta}{2\vartheta}{\cal I}_2+\frac{\vartheta}{\bar \vartheta}\left[\ln \vartheta+\frac{1-\vartheta^{-2\gamma}}{2}\left(\chi(n,\gamma)-2\ln \bar \vartheta\right)\right] \; ,
\end{equation}
and
\begin{equation}
\stepcounter{appcnt}
\label{I3}
 {\cal I}_3=\frac{\bar \vartheta}{2\vartheta}{\cal I}_2-\frac{\vartheta}{\bar \vartheta}\left[\ln \vartheta+\frac{1-\vartheta^{-2\gamma}}{2}\left(\chi(n,\gamma)-2\ln \bar \vartheta\right)\right] \; ,
\end{equation}
with ${}_2F_1$ the Gauss hypergeometric function.
The \emph{plus-prescription} in Eqs.~\eqref{Cgg_hadron} and~\eqref{Cqq_hadron} act as
\begin{equation}
\label{plus-prescription}
\stepcounter{appcnt}
\int^1_\zeta \drv \zeta \frac{f(\zeta)}{(1-\zeta)_+}
=\int^1_\zeta \drv \zeta \frac{f(\zeta)-f(1)}{(1-\zeta)}
-\int^\zeta_0 \drv \zeta \frac{f(1)}{(1-\zeta)}\; 
\end{equation}
on any function $f(\zeta)$ regular at $\zeta=1$.

\end{appendices}

%\bibliographystyle{cas-model2-names}
\bibliographystyle{elsarticle-num}

%-----------------------------------------
\bibliography{references}
%-----------------------------------------

\end{document}